\documentclass[aps,prb,twocolumn,showpacs,amsmath,amssymb,footinbib,longbibliography]{revtex4-2}
\usepackage[percent]{overpic} 
\usepackage{graphicx}  
\usepackage{mathtools}
\usepackage{bm}         
\usepackage[dvipsnames]{xcolor} 
\usepackage{tikz}       
\usepackage{mathrsfs}
\usepackage{subfigure, psfrag}
\usepackage[colorlinks=true,urlcolor=blue,citecolor=blue,linkcolor=blue]{hyperref}
\hypersetup{breaklinks=true}
\begin{document}
\title{Kondo transport in an anisotropic two-dimensional electron gas with quadratic momentum-dependent spin splitting}
\author{M. R. Parymuda}
\email{mparymuda@icmp.lviv.ua}
\affiliation{Yukhnovskii Institute for Condensed Matter Physics of the National Academy of Sciences of Ukraine, Svientsitskii Street 1, 79011 Lviv, Ukraine}
\begin{abstract}
We investigate the transport properties of an anisotropic two-dimensional electron gas with a quadratic spin texture, described by a low-energy effective $\bm k\cdot \bm p$ model, in the presence of $S=1/2$ Kondo impurities. We develop a Green's function framework that captures the interplay between the spin texture and the Kondo scattering for calculating transport properties in such systems. Using this framework together with numerical renormalization group (NRG) calculations, we evaluate the longitudinal resistivity components $\rho^{xx}(T)$ and $\rho^{yy}(T)$ and analyze their temperature dependence. We further determine the Kondo temperature $T_K(\alpha)$ and the impurity screening function $p(\alpha)$ and demonstrate that they decrease strongly with increasing quadratic spin-splitting coupling $\alpha$. 

\end{abstract}
\maketitle

\section{Introduction}
\label{sec:1}
The Kondo effect has long been a fundamental topic in condensed-matter physics because of its well-known low-temperature anomaly in electronic transport. It is a many-body phenomenon arising from the antiferromagnetic exchange interaction between conduction electrons and localized magnetic moments. In dilute magnetic alloys, this interaction enhances spin-flip scattering at low temperatures, producing a resistivity minimum, below which the resistivity rises logarithmically with decreasing temperature \cite{kondo1964,yanagisawa2023,kondo1970,kouwenhoven2001}.

The Kondo effect also appears in tunable mesoscopic metals. In quantum dots, a localized spin on the dot couples antiferromagnetically to electrons in the leads via higher-order tunneling processes, producing a Kondo singularity in the density of states at the Fermi level and an enhanced conductance \cite{liu2011,cronenwett1998,pustilnik2004,busz2022}.

Previous studies of Kondo transport have largely focused on spin-degenerate conduction bands and systems characterized by linear Rashba \cite{bychkov1984} and Dresselhaus \cite{dresselhaus1955} spin splitting. The recent identification of quadratic spin-texture materials and altermagnets opens a new direction for studying the Kondo effect in systems exhibiting quadratic spin splitting. Such momentum-dependent spin splitting has been reported in several compounds, including BaSbPt \cite{liu2024}, MnTe$_2$ \cite{zhu2024}, and AgTlP$_2$Se$_6$ \cite{zhou2026}. These materials motivate a systematic investigation of how magnetic impurities affect transport in systems exhibiting quadratic spin splitting. 

In such systems, Kondo transport is governed by the competition between Kondo correlations arising from the exchange interaction $J_{sd}$ and the quadratic spin-splitting coupling $\alpha$. The exchange interaction generates spin-flip scattering responsible for impurity screening, whereas $\alpha$ separates the conduction band into two helicity branches with momentum-dependent spin orientations and thereby modifies the electronic states participating in these processes. Clarifying how this interplay modifies Kondo screening and transport is therefore central to understanding the role of localized moments in materials with quadratic spin splitting.

Despite growing interest in systems with momentum-dependent spin splitting, their transport properties have received comparatively little theoretical attention, particularly in the presence of Kondo scattering from magnetic impurities. We address this gap by calculating the temperature-dependent Kondo resistivity using a self-consistent Green's function method \cite{nagaoka1965,yanagisawa2012,yanagisawa2015} and numerical renormalization group (NRG) calculations \cite{zitko2009,zitko2009_1,bulla2008}.

The paper is organized as follows. In Sec.~\ref{sec:2}, we introduce a low-energy effective $\bm k\cdot \bm p$ model describing a quadratic spin texture, together with the Kondo model for magnetic impurities. In Sec.~\ref{sec:3}, we develop a Green's function framework to describe the interaction between magnetic impurities and the conduction electrons, governed by the exchange coupling $J_{sd}$, in the presence of the quadratic spin-splitting coupling $\alpha$. In Sec.~\ref{sec:4}, we determine how the Kondo temperature $T_K(\alpha)$ and the impurity screening function $p(\alpha)$ depend on the quadratic spin-splitting coupling $\alpha$, using both the Green's function formalism and the NRG framework. In Sec.~\ref{sec:5}, we employ the two methods to investigate how $\alpha$ and the effective masses $m_x$ and $m_y$ influence the temperature dependence of the resistivity. In Sec.~\ref{sec:6}, we discuss our results. In Sec.~\ref{sec:7}, we summarize our findings. 

Throughout this paper, we use $\hbar=k_B=1$ and restore these constants in the final expressions.

\vspace{5mm}

\section{Model}
\label{sec:2}
We begin by introducing the Hamiltonian of the system under consideration
\begin{equation}\mathcal H=\mathcal H_{0}+\mathcal H_{sd}.
\label{eq:1}
\end{equation}

The Hamiltonian of the two-dimensional electron gas, including the quadratic spin-splitting term arising from the $\bm k\cdot \bm p$ expansion, is given by \cite{liu2024}
\begin{equation*}\mathcal H_0=\sum_{\bm k}\psi^\dagger_{\bm k}\Big(\xi_{\bm k}\sigma_0+\alpha \left((k_x^2-k_y^2)\sigma_x-2k_xk_y\sigma_y\right)\Big)\psi_{\bm k}
\end{equation*}	
\begin{equation}
=\sum_{\bm k}\Big[\xi_{\bm k}(C^\dagger_{\bm k\uparrow}C_{\bm k\uparrow}+C^\dagger_{\bm k\downarrow}C_{\bm k\downarrow})+\alpha k_+^2C^\dagger_{\bm k\uparrow}C_{\bm k\downarrow}+\alpha k_-^2 C^\dagger_{\bm k\downarrow}C_{\bm k\uparrow}\Big].
\label{eq:2}
\end{equation}

In Hamiltonian~(\ref{eq:2}), $ \psi^\dagger_{\bm k}=\begin{pmatrix}
C^\dagger_{\bm k\uparrow} & C^\dagger_{\bm k\downarrow}	
\end{pmatrix}$ is the spinor creation operator, $k_\pm=k_x\pm ik_y$, and $\sigma_x$ and $\sigma_y$ denote the Pauli matrices. The anisotropic spectrum is $\xi_{\bm k}=k_x^2/2m_x+k_y^2/2m_y-\mu$, where $\mu$ is the chemical potential. The parameter $\alpha$ denotes the strength of the quadratic spin-splitting coupling. The operators $C^\dagger_{\bm k\sigma}$ and $C_{\bm k\sigma}$ create and annihilate an electron with momentum $\bm k$ and spin $\sigma$, respectively.

\begin{figure}[]
  \centering
  \subfigure[]{\includegraphics[width=1.01\linewidth]{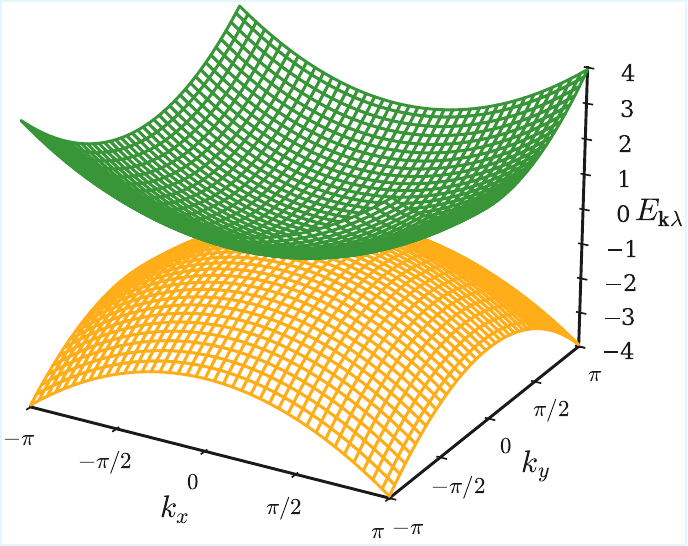}}
  \subfigure[]{\includegraphics[width=0.8\linewidth]{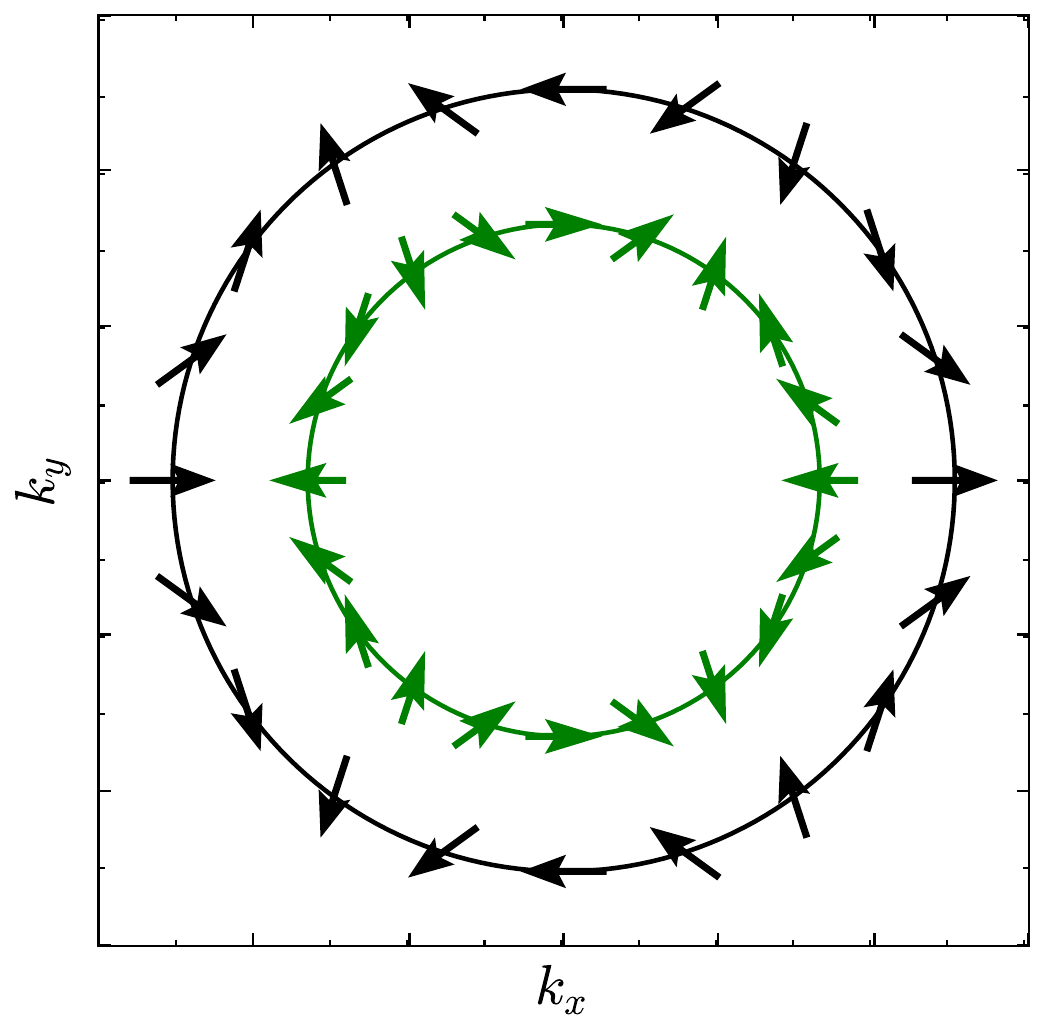}}
 \caption{(a) Energy spectrum of the quadratic spin-splitting model. (b) Schematic illustration of the momentum-space spin texture generated by the spin-orbit field $\bm g_\alpha(\bm k)$. The arrows indicate the spin polarization in the two helicity bands, aligned parallel or antiparallel to $\bm g_{\alpha}(\bm k)$.}
  \label{fig:1}
  \end{figure}

The energy spectrum of Hamiltonian~(\ref{eq:2}) is given by
\begin{equation}\xi_{\bm k\lambda}=\xi_{\bm k}+E_{\bm k\lambda},
\label{eq:3}	
\end{equation}
where $E_{\bm k\lambda}=\lambda \alpha k^2 $, and $\lambda=\pm 1$ labels the upper ($+$) and lower ($-$) bands, respectively. The spectrum $E_{\bm k\lambda}$ are shown in Fig.~\ref{fig:1}(a).

The quadratic spin-mixing term in Hamiltonian~(\ref{eq:2}) can be interpreted as a momentum-dependent effective spin-orbit field \cite{manchon2015}

\begin{equation}
\mathcal H^{split}_{\bm k}=\bm g_\alpha(\bm k)\cdot \bm \sigma,
\label{eq:4}
\end{equation}
where $\bm g_\alpha(\bm k)=\alpha k^2(\cos 2\varphi,\;-\sin2\varphi)$ denotes the corresponding effective field, cf. Ref.~\cite{neilo2022}. This field lies entirely in the $x$-$y$ plane, with magnitude $\alpha k^2$,  while its orientation is set by the polar angle $\varphi$. As $\varphi$ varies over a full $2\pi$, the direction of $\bm  g_\alpha(\bm k)$ winds twice around the origin, as illustrated in Fig.~\ref{fig:1}(b).

The exchange interaction between a Kondo impurity and conduction electrons is given by \cite{nagaoka1965}
$$\mathcal H_{sd}=-\frac{J_{sd}}{2N_s}\sum_{\bm k\bm k'}\big[(C^\dagger_{\bm k'\uparrow}C_{\bm k\uparrow}-C^\dagger_{\bm k'\downarrow}C_{\bm k\downarrow})S^z$$
\begin{equation}+C^\dagger_{\bm k'\downarrow}C_{\bm k\uparrow}S^++C^\dagger_{\bm k'\uparrow}C_{\bm k\downarrow}S^-\big].
\label{eq:5}	
\end{equation}

Here, $J_{sd}$ denotes the antiferromagnetic exchange coupling ($J_{sd}<0$) between the localized impurity spin and the conduction electrons. The operators $S^\pm$ and $S^z$ denote the transverse and longitudinal components of the impurity spin operator, corresponding to the $x$-$y$ plane and the $z$ axis, respectively.

\vspace{15mm}

\section{Green's function}
\label{sec:3}
We define the following Green's functions
\begin{equation}\mathcal G_{\bm k\bm k'}(\tau)=-\langle T_\tau C_{\bm k\uparrow}(\tau)C^\dagger_{\bm k'\uparrow}(0)\rangle=\langle\!\langle C_{\bm k\uparrow}|C^\dagger_{\bm k'\uparrow}\rangle\!\rangle,
\label{eq:6}	
\end{equation}
\begin{equation}\mathcal F_{\bm k\bm k'}(\tau)=-\langle T_\tau C_{\bm k\downarrow}(\tau)C^\dagger_{\bm k'\uparrow }(0)\rangle=\langle\!\langle C_{\bm k\downarrow}|C^\dagger_{\bm k'\uparrow}\rangle\!\rangle.
\label{eq:7}	
\end{equation}

Their Fourier transforms are defined as
\begin{equation}
\langle\!\langle C_{\bm k\sigma}|C^\dagger_{\bm k'\sigma'}\rangle\!\rangle_\tau =\frac{1}{\beta}\sum_{\omega_n}e^{-i\omega_n\tau} \langle\!\langle C_{\bm k\sigma}|C^\dagger_{\bm k'\sigma'}\rangle\!\rangle_{i\omega_n},
\label{eq:8}
\end{equation}
where $\omega_n=\pi(2n+1)/\beta$ are the fermionic Matsubara frequencies and $\beta=1/T$.

We begin with the general equation of motion (EOM) for the Green's function
\begin{equation}i\omega_n \langle\!\langle A_{\bm k\sigma}|B_{\bm k'\sigma'}\rangle\!\rangle=\langle \{A_{\bm k\sigma},B_{\bm k'\sigma'}\}\rangle+\langle\!\langle \dot A_{\bm k\sigma}|B_{\bm k'\sigma'}\rangle\!\rangle,
\label{eq:9}
\end{equation}
where $\dot A_{\bm k\sigma}=[A_{\bm k\sigma},\mathcal H]$, and  $\sigma,\;\sigma'=\uparrow, \downarrow$. 

Applying Eq.~(\ref{eq:9}) to the Green's functions $\mathcal G_{\bm k\bm k'}$ and $\mathcal F_{\bm k \bm k'}$, we obtain the following coupled EOMs

\begin{equation}
(i\omega_n-\xi_{\bm k})\mathcal G_{\bm k\bm k'}=\delta_{\bm k\bm k'}+\alpha k_+^2\mathcal F_{\bm k\bm k'}-\frac{J_{sd}}{2 N_s}\sum_{\bm q}\Gamma_{\bm q\bm k'},
\label{eq:10}
\end{equation}
\vspace{-3mm}
\begin{equation}
(i\omega_n-\xi_{\bm k})\mathcal F_{\bm k\bm k'}=\alpha k_-^2\mathcal G_{\bm k\bm k'}+\frac{J_{sd}}{2 N_s}\sum_{\bm q}\Delta_{\bm q\bm k'}.
\label{eq:11}	
\end{equation}
In Eqs.~(\ref{eq:10}) and (\ref{eq:11}), we introduced the following functions
\begin{equation}
\Gamma_{\bm k\bm k'}=\langle\!\langle C_{\bm k\uparrow}S^z|C^\dagger_{\bm k'\uparrow}\rangle\!\rangle+\langle\!\langle C_{\bm k\downarrow}S^-|C^\dagger_{\bm k'\uparrow}\rangle\!\rangle,
\label{eq:12}
\end{equation}
and
\begin{equation}
\;\;\;\;\;\;\Delta_{\bm k\bm k'}=\langle\!\langle C_{\bm k\downarrow}S^z|C^\dagger_{\bm k'\uparrow}\rangle\!\rangle-\langle\!\langle C_{\bm k\uparrow}S^+|C^\dagger_{\bm k'\uparrow}\rangle\!\rangle.
\label{eq:13}
\end{equation}

We further derive the EOMs for the auxiliary Green's functions defined in Eqs.~(\ref{eq:12}) and (\ref{eq:13}) and substitute the resulting expressions into Eqs.~(\ref{eq:10}) and (\ref{eq:11}). Since the intermediate calculations are lengthy, the details are presented in Appendices~\ref{appendix:A}, \ref{appendix:B}, and \ref{appendix:C}. Here, we give only the final expressions for the dressed single-particle Green's functions

\begin{widetext}
{\small
\begin{subequations}
\begin{equation}
\mathcal G_{\bm k\bm k'}(i\omega_n)=g_{\bm k}(i\omega_n)\delta_{\bm k\bm k'}+\frac{\mathcal U_{\bm k}(i\omega_n)(1-\mathcal B(i\omega_n))+\mathcal R_{\bm k}(i\omega_n)\mathcal I(i\omega_n)}{(1-\mathcal U(i\omega_n))(1-\mathcal B(i\omega_n))-\mathcal R(i\omega_n)\mathcal I(i\omega_n)}g_{\bm k'}(i\omega_n)
+\frac{\mathcal U_{\bm k}(i\omega_n)\mathcal R(i\omega_n)+\mathcal R_{\bm k}(i\omega_n)(1-\mathcal U(i\omega_n))}{(1-\mathcal U(i\omega_n))(1-\mathcal B(i\omega_n))-\mathcal R(i\omega_n)\mathcal I(i\omega_n)}h_{\bm k'-}(i\omega_n),
\label{eq:14a}
\end{equation}
and
\begin{equation}
\mathcal F_{\bm k\bm k'}(i\omega_n)=h_{\bm k-}(i\omega_n)\delta_{\bm k\bm k'}+\frac{\mathcal I_{\bm k}(i\omega_n)\mathcal (1-\mathcal B(i\omega_n))+\mathcal B_{\bm k}(i\omega_n)\mathcal I(i\omega_n)}{(1-\mathcal U(i\omega_n))(1-\mathcal B(i\omega_n))-\mathcal R(i\omega_n)\mathcal I(i\omega_n)}g_{\bm k'}(i\omega_n)
+ \frac{\mathcal I_{\bm k}(i\omega_n)\mathcal R(i\omega_n)+\mathcal B_{\bm k}(i\omega_n)(1-\mathcal U(i\omega_n))}{(1-\mathcal U(i\omega_n))(1-\mathcal B(i\omega_n))-\mathcal R(i\omega_n)\mathcal I(i\omega_n)}h_{\bm k'-}(i\omega_n),
\label{eq:14b}
\end{equation}
\end{subequations}}
\end{widetext}
where 
\begin{equation*}
\mathcal U(i\omega_n)=\sum_{\bm k}\mathcal U_{\bm k}(i\omega_n),\qquad \mathcal R(i\omega_n)=\sum_{\bm k}\mathcal R_{\bm k}(i\omega_n),
\end{equation*}
\begin{equation}
\mathcal I(i\omega_n)=\sum_{\bm k}\mathcal I_{\bm k}(i\omega_n),\qquad \mathcal B(i\omega_n)=\sum_{\bm k}\mathcal B_{\bm k}(i\omega_n).
\label{eq:15}
 \end{equation}
In Eqs.~(\ref{eq:14a}) and (\ref{eq:14b}), $g_{\bm k}(i\omega_n)$ and $h_{\bm k\pm}(i\omega_n)$ denote the diagonal and off-diagonal components of the noninteracting Green's function associated with Hamiltonian~(\ref{eq:2}), respectively. Their explicit forms are
\begin{equation}g_{\bm k}(i\omega_n)=\frac{i\omega_n-\xi_{\bm k}}{(i\omega_n-\xi_{\bm k})^2-\alpha^2k^4},
\label{eq:16}
\end{equation}
and
\begin{equation}h_{\bm k\pm}(i\omega_n)=\frac{\alpha k_\pm^2}{(i\omega_n-\xi_{\bm k})^2-\alpha^2k^4}.
\label{eq:17}
\end{equation}

\section{Kondo screening}
\label{sec:4}
\subsection{Evolution of the Kondo temperature: Green's function approach}

The Kondo temperature is determined from the poles of the Green's function. Performing the analytic continuation from Matsubara frequencies to real frequencies, we obtain the condition ${\rm Re}\mathcal E(\omega+i\delta)=0$, which yields  

\begin{equation}
1+\left(\frac{J_{sd}}{N_s}\right){\rm Re}\sum_{\bm k}\frac{\omega-\xi_{\bm k}}{(\omega-\xi_{\bm k})^2-\alpha^2(k_x^2+k_y^2)^2}\left(n_{\bm k}-\frac{1}{2}\right)=0.
\label{eq:18}
\end{equation}
We construct a matrix composed of the eigenvectors of the Hamiltonian $\mathcal H_0$ as
\begin{equation}
\bm U_{\bm k}=\begin{pmatrix}
    u_{\bm k+}& u_{\bm k-}
\end{pmatrix}=\frac{1}{\sqrt{2}}\begin{pmatrix}
	1 & 1\\
	 e^{-i2\varphi_{\bm k}} & -e^{-i2\varphi_{\bm k}}
\end{pmatrix}.
\label{eq:19}
\end{equation}
The matrix $\bm U_{\bm k}$ is determined in such a manner that it diagonalizes the Hamiltonian~(\ref{eq:2}), i.e., $\bm U^\dagger_{\bm k}\mathcal H_{0\bm k} \bm U_{\bm k}=diag(\xi_{\bm k+},\;\xi_{\bm k-})$. The fermionic operators in the original basis are then related to those in the diagonal (eigenmode) basis by $C_{\bm k\mu}=\sum_{\nu=1,2}U^{\mu\nu}_{\bm k}\gamma_{\bm k\nu}$ and $C^\dagger_{\bm k\mu}=\sum_{\nu=1,2}\left(U^{\mu\nu}_{\bm k}\right)^*\gamma^\dagger_{\bm k\nu}$, where $\gamma_{\bm k\nu}$ and $\gamma^\dagger_{\bm k\nu}$ are the annihilation and creation operators in the eigenmode basis, respectively. Accordingly, Hamiltonian~(\ref{eq:2}) can be recast into the form $\mathcal H_0=\sum_{\bm k}\sum_{\lambda=\pm}\xi_{\bm k\lambda}\gamma^\dagger_{\bm k\lambda}\gamma_{\bm k\lambda}$.

Using the unitary matrix $\bm U_{\bm k}$, we evaluate the occupation number to zeroth order in the exchange interaction $J_{sd}$ \cite{yanagisawa2012,yanagisawa2015} 
\begin{equation*}
n^{(0)}_{\bm k}=\langle C^\dagger_{\bm k\uparrow}C_{\bm k\uparrow}\rangle=\langle C^\dagger_{\bm k\downarrow}C_{\bm k\downarrow}\rangle=\sum_{\lambda=\pm}\left|U^{+\lambda}_{ \bm k }\right|^2n_F(\xi_{\bm k\lambda})
\end{equation*}
\vspace{-5mm}
\begin{equation}=\frac{1}{2}\left(n_F(\xi_{\bm k}-\alpha k^2)+n_F(\xi_{\bm k}+\alpha k^2)\right),
\label{eq:20}
\end{equation}
where $n_F(\xi_{\bm k\lambda})=\langle \gamma^\dagger_{\bm k\lambda}\gamma_{\bm k\lambda}\rangle=1/({\rm exp}(\beta\xi_{\bm k\lambda})+1)$ denotes the Fermi distribution function.

By substituting Eq.~(\ref{eq:20}) into Eq.~(\ref{eq:18}), we obtain 
\begin{equation}
{\rm Re}\int_{-\mu}^{D-\mu} d\xi  \sum_{\lambda,\lambda'=\pm}\frac{n_F(\xi_{\bm k\lambda'})-1}{\omega-\xi_{\bm k\lambda}+i0^+}=-\frac{4}{J_{sd}\nu_{2D}},
\label{eq:21}
\end{equation}
where $\nu_{2D}=\sqrt{m_xm_y}/2\pi.$

In deriving Eq.~(\ref{eq:21}), we assume the weak spin-splitting regime $\alpha m_{x,y} \ll 1$, so that the term $\alpha k^2$ remains small compared with the other energy scales appearing in the denominators. Therefore, we replace $\alpha k^2$ by its Fermi average $\alpha \langle k^2\rangle_{FS}$, where $\langle k^2\rangle_{FS}=\langle k^2 \rangle
=\mu(m_x+m_y)$, and $\langle \hdots\rangle_{FS}=\int d^2k(\hdots)\delta(\xi_{\bm k})/\int d^2k \delta(\xi_{\bm k})$.

For the evaluation of Eq.~(\ref{eq:21}), we introduce the following integrals
\begin{equation*}
{\rm Re}\int_{-\mu}^{D-\mu} d\xi \frac{n_F\left(\xi\pm \alpha \langle k^2\rangle\right)-\frac{1}{2}}{\omega-\xi\pm\alpha \langle k^2\rangle+i0^+}
\end{equation*}
\vspace{-5mm}
\begin{equation}=\ln\left(\frac{ \sqrt{\mu(D-\mu)}}{2\pi T}\right)-{\rm Re}\psi\left(\frac{1}{2}-\frac{i(\omega\mp 2\alpha\langle k^2\rangle)}{2\pi T}\right),
\label{eq:22}
\end{equation}
and
\begin{equation*}
{\rm Re}\int_{-\mu}^{D-\mu} d\xi \frac{n_F\left(\xi \mp \alpha \langle k^2\rangle\right)-\frac{1}{2}}{\omega-\xi\pm \alpha \langle k^2\rangle+i0^+}
\end{equation*}
\vspace{-5mm}
\begin{equation}=\ln\left(\frac{\sqrt{ \mu(D-\mu)}}{2\pi T}\right)-{\rm Re}\psi\left(\frac{1}{2}-\frac{i\omega}{2\pi T}\right),
\label{eq:23}
\end{equation}
where $\psi(z)$ is the digamma function.

These integrals, see Eqs.~(\ref{eq:22}) and (\ref{eq:23}), are evaluated as follows. First, we rewrite the Fermi function as $n_F(\xi)-1/2=-1/2 \tanh\left(\beta\xi/2\right)$, and then decompose the kernel according to $1/(\omega-\xi)=-1/\xi+\omega/[\xi(\omega-\xi)]$. The first integral encountered is $\int_{-\mu}^{D-\mu}d\xi \tanh\left(\beta \xi/2\right)/\xi=2\ln\left(2e^\gamma \sqrt{\mu(D-\mu)}/\pi T\right)$, see Ref.~\cite{bruus2004}, while the second integral reads $\int_{-\mu}^{D-\mu}d\xi \tanh\left(\beta\xi/2\right)/[\xi(\xi-\omega)]=2\left(\psi(1/2)-\psi(1/2-i\omega/2\pi T)\right)/\omega$, where we have assumed $T,\,|\omega|,\, \alpha \langle k^2\rangle\ll \mu, \; D-\mu$. The second integral is evaluated using the residue theorem, see Ref.~\cite{mattuck1992}. 

We therefore arrive at the expression
\begin{equation*}
\frac{4}{|J_{sd}|\nu_{2D}}=4\ln \left(\frac{ \sqrt{\mu(D-\mu)}}{2\pi T_K}\right)-{\rm Re}\Bigg[2\psi\Bigg(\frac{1}{2}-\frac{i\omega}{2\pi T}\Bigg)
\end{equation*}
\begin{equation}
+\psi\left(\frac{1}{2}-\frac{i\left(\omega-2\alpha\langle k^2\rangle\right)}{2\pi T_K}\right)+\psi\left(\frac{1}{2}-\frac{i\left(\omega+2\alpha\langle k^2\rangle\right)}{2\pi T_K}\right)\Bigg].
\label{eq:24}
\end{equation}

Indeed, setting $\alpha=0$ and $\omega=0$ in Eq.~(\ref{eq:22}), we recover the standard expression for the Kondo temperature in a spin-degenerate electron gas
\begin{equation}
T^{(0)}_K=\frac{2e^\gamma\sqrt{\mu (D-\mu)}}{\pi}{\rm exp}\left(-\frac{1}{|J_{sd}|\nu_{2D}}\right),
\label{eq:25}
\end{equation}
where $\gamma=0.5772$ is the Euler constant.

\begin{figure}[]
  \centering
  {\includegraphics[width=0.9\linewidth]{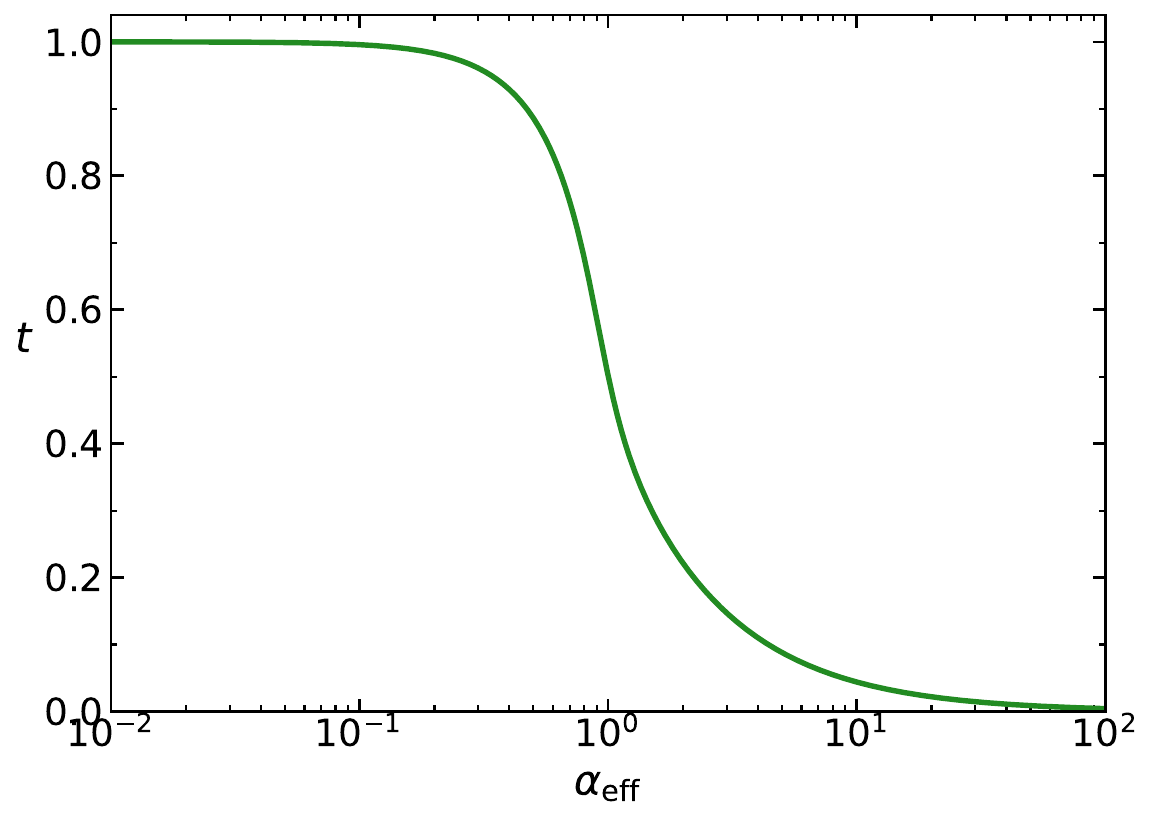}}
\caption{Dependence of the normalized Kondo temperature $t$ on the effective spin-splitting coupling $\alpha_{eff}$.} 
\label{fig:2}
  \end{figure}

For the numerical  evaluation, we substract Eq.~(\ref{eq:22}) evaluated at $\alpha=0$ from the same equation at finite $\alpha$ and normalize the resulting expression by the bare Kondo temperature $T_K^{(0)}$. At $\omega=0$, the resulting expression takes the form 
\begin{equation}t={\rm exp}\left(-\frac{1}{2}\left[{\rm Re}\psi\left(\frac{1}{2}+\frac{i\alpha_{eff}}{\pi t}\right)-\psi\left(\frac{1}{2}\right)\right]\right),
\label{eq:26}
\end{equation}
where
\begin{equation}
t=T_K(\alpha)/T_K^{(0)},\qquad \alpha_{eff}=\alpha \langle k^2\rangle/ T^{(0)}_K,
\label{eq:27}
\end{equation}
The graphical solution of the self-consistency equation $f(t)=t$ is shown in Fig.~\ref{fig:2}. In the limit $\alpha_{eff}\to 0$, this solution continuously approaches $t \to 1$, which is consistent with the standard two-dimensional electron gas. Indeed, expanding Eq.~(\ref{eq:26}) for small $\alpha_{eff}$, we obtain $t\simeq 1-\frac{7\zeta(3)}{2\pi^2} \alpha_{eff}^2$. Thus, even weak quadratic spin splitting reduces the Kondo scale.
At large $\alpha_{eff}$, the solution behaves as $t\simeq\pi e^{-\gamma}/4\alpha_{eff}$, demonstrating that $T_K(\alpha)\to 0$ as the quadratic spin-splitting coupling $\alpha$ increases.

The suppression of $T_K(\alpha)$ is a consequence of the competition between the Kondo effect and the quadratic spin splitting. Kondo correlations arise from spin-flip exchange processes between the impurity and the conduction electrons. As $\alpha$ increases, the splitting energy $\Delta_{split}\simeq 2\alpha \langle k^2\rangle_{FS}$, see Eq.~(\ref{eq:24}), between the helicity branches becomes larger, making inter-helicity spin-flip processes less favorable and thereby suppressing Kondo screening. The reduction of $T_K(\alpha)$ shown in Fig.~\ref{fig:2} is gradual because increasing $\alpha$ weakens the inter-helicity spin-flip contributions, whereas the intra-helicity contributions remain finite and continue to participate in Kondo screening, see Fig.~\ref{fig:1}(b).

\subsection{Impurity screening function $p(\alpha)$: NRG calculation}
\label{sec:4B}

We now investigate the influence of the quadratic spin-splitting coupling $\alpha$ on the impurity screening function $p(\alpha)$ using the NRG within the Ljubljana framework \cite{zitko2009,zitko2009_1,bulla2008}. We define this function as $p(\alpha)=1-|\langle S^x\rangle|/S$, which can be obtained directly from the NRG calculations and characterizes the weakening of impurity screening with increasing $\alpha$.

\begin{figure}[!b]
  \centering

{\includegraphics[width=1\linewidth]{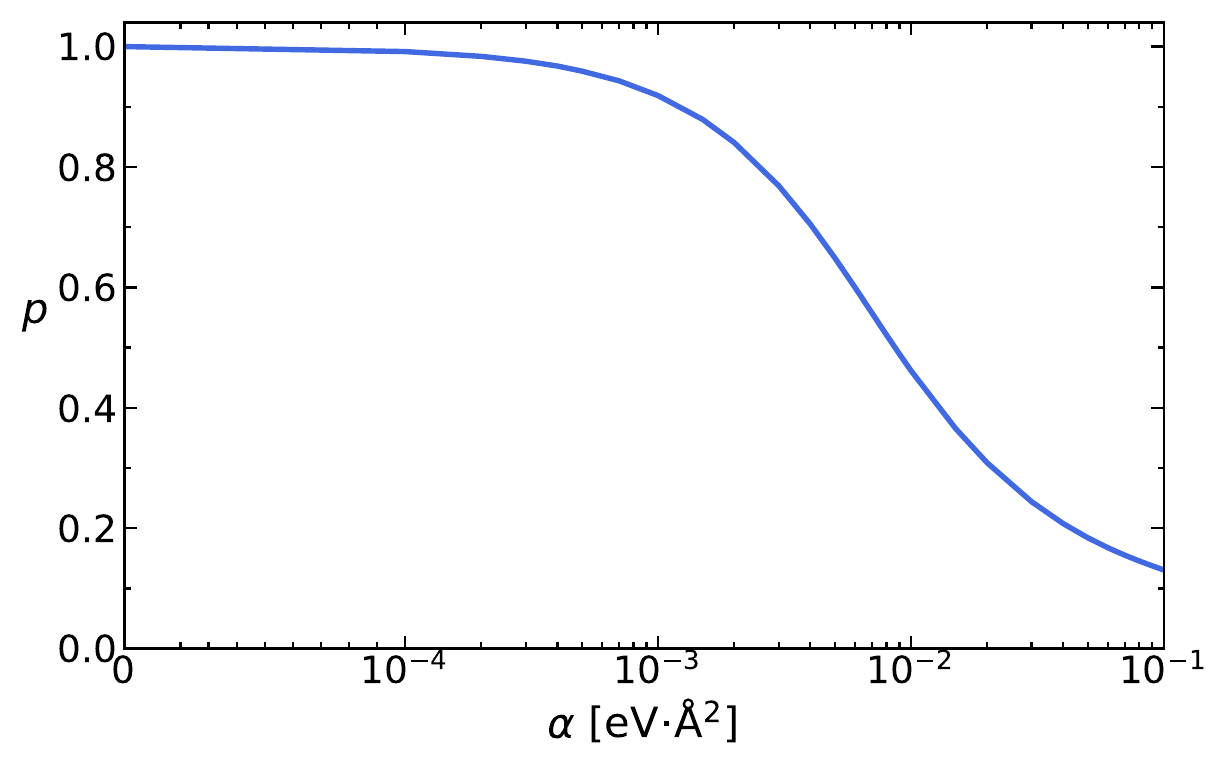}}
\caption{Dependence of the impurity screening function $p(\alpha)$ on the quadratic spin-splitting coupling $\alpha$, calculated using the Ljubljana NRG framework \cite{zitko2009,zitko2009_1,bulla2008}. The parameters are $J_{sd}=0.7\;{\rm eV},\;\mu=0.25\; {\rm eV}$, $D=0.5\; {\rm eV}$, $n_{imp}=10^{-2}$, $m_x=1.8$, and $\;m_y=1$. The effective masses $m_x$ and $m_y$ are expressed in units of the free-electron mass $m_e$.}
\label{fig:3}
  \end{figure}

For the NRG calculation of the impurity screening function $p(\alpha)$, we first define the local density of states (DOS) of the two-dimensional electron gas resolved with respect to the $x$-component of spin as 
\begin{equation}
\nu_s(\omega)=-\frac{1}{\pi}{\rm Im}\int \frac{d^2k}{(2\pi)^2}  {\rm Tr}\left(\hat{P}_s \bm g_{\bm k}^R(\omega)\right).
\label{eq:28}
\end{equation}
Here, $\hat{P}_s=(1+s\sigma_x)/2$ is the spin-projection operator, while the $2\times 2$ retarded Green's function matrix corresponding to Hamiltonian~(\ref{eq:2}) is given by  
\begin{equation}\bm g^R_{\bm k}(\omega)=\begin{pmatrix}
    g_{\bm k}^R(\omega)& h^R_{\bm k+}(\omega)\\\\
    h^R_{\bm k-}(\omega)& g^R_{\bm k}(\omega)
\end{pmatrix}.
\label{eq:29}
\end{equation}

Inserting Eq.~(\ref{eq:29}) into Eq.~(\ref{eq:28}), we obtain
\begin{equation}\nu_s(\omega)=\sum_{\lambda=\pm}\int \frac{d^2k}{(2\pi)^2}\frac{1+s\lambda \cos(2\varphi_{\bm k})}{2}\delta(\omega-\xi_{\bm k\lambda})
\label{eq:30}
\end{equation}

We determine the normalization of the DOS for each spin component as
\begin{equation}N_s=\int_{-W}^{W}\nu_s(\omega)d\omega,
\label{eq:31}
\end{equation}
where $W=D-\mu+2\alpha Dm_x$ is the upper boundary of the physical energy interval. Throughout this paper, we consider $m_x>m_y$ and $D> 2\mu/(1+2\alpha m_x)$, which ensures that $W> \mu$. For the NRG calculations, the physical energy interval $\omega \in [-\mu,\;\; W]$ is extended to the symmetric interval $\omega \in[-W,\; W]$. In the added energy interval, the DOS is taken to have a negligibly small positive value, so that the symmetric extension has a negligible effect on its normalization. Consequently, the DOS has the same normalization for both spin components $N_+=N_-=N=D\nu_{2D}$.

We next introduce the dimensionless frequency $\widetilde{\omega}=\omega/W$, with $-1\leq \widetilde{\omega}\leq 1$, and define the dimensionless DOS as 
\begin{equation}\widetilde{\nu}_s(\widetilde{\omega})=\frac{W}{N}\nu_s(\widetilde{\omega}W),\qquad\int_{-1}^{1}\widetilde{\nu}_s(\widetilde{\omega})d\widetilde{\omega}=1
\label{eq:32}
\end{equation}
The normalization of the DOS requires a corresponding rescaling of the exchange coupling used in the NRG calculations as follows
\begin{equation}J^{NRG}_{sd}\widetilde{\nu}_s(\widetilde{\omega})=J_{sd}\nu_s(\omega),\qquad J_{sd}^{NRG}=J_{sd}\frac{N}{W}.
\label{eq:33}
\end{equation}

For each value of $\alpha$, the NRG calculations are performed with the discretization parameter $\Lambda=2$ and $z$-averaging over $N_z=4$ discretization meshes defined by $z_i=(i-1/2)/N_z$ with $i=1,\hdots,N_z$. At least 6000 states are retained at each iteration. The impurity polarization in the zero-temperature limit is averaged over the discretization meshes as follows 
\begin{equation}\langle S^x\rangle=\frac{1}{N_z}\sum_i \langle S^x\rangle_{z_i,\; T\to 0} .
\label{eq:34}
\end{equation}
The resulting polarization is then used to evaluate the impurity screening function $p(\alpha)$.

We next examine the behavior of the $p(\alpha)$, which is displayed in Fig.~\ref{fig:3}. At $\alpha=0$, the conduction band is spin degenerate, so that the impurity spin remains unpolarized $\langle S^x\rangle=0$, and the impurity screening function takes its maximum value $p(0)=1$. A finite $\alpha$ alone, however, is insufficient to polarize the impurity. In the isotropic limit $m_x=m_y$, the impurity remains unpolarized for any $\alpha$, so that $\langle S^x\rangle=0$ and $p(\alpha)=1$. By contrast, when $m_x\neq m_y$, the impurity screening function $p(\alpha)$ decreases monotonically with increasing $\alpha$, reflecting the development of a finite impurity polarization $\langle S^x\rangle\neq0$.

The microscopic origin of this behavior lies in the effective field $h_{eff}$ generated through the $s$-$d$ exchange interaction and given by $h_{eff}=\frac{J_{sd}}{2}\int d\omega n_F(\omega)[\nu_-(\omega)-\nu_+(\omega)]$. This field becomes nonzero when a finite $\alpha$, together with mass anisotropy, produces unequal spin-resolved DOS $\nu_-\neq \nu_{+}$. This relation clarifies the physical origin of the impurity polarization, explaining why the impurity remains unpolarized in the isotropic system even at finite $\alpha$, whereas a finite impurity polarization develops when both finite $\alpha$ and mass anisotropy are present.


\section{Resistivity}
\label{sec:5}
In this section, we present the calculation of the resistivity $\rho^{\gamma\gamma}=1/\sigma^{\gamma\gamma}$ for an anisotropic two-dimensional electron gas with quadratic spin splitting, described by Hamiltonian~(\ref{eq:1}), in the dilute-impurity limit. We evaluate the longitudinal resistivity components $\rho^{xx}$ and $\rho^{yy}$ using the Green's function method and NRG calculation.

The dc charge conductivity along the $\gamma$ direction is defined as
\begin{equation}
\sigma^{\gamma\gamma}(T)=\frac{e^2}{N_s}\sum_{\bm k}\sum_{\lambda=\pm} v_{\bm k\lambda,\gamma}^2\tau_{\bm k\lambda}(\xi_{\bm k\lambda},T)\left(-\frac{\partial n_F(\xi_{\bm k\lambda})}{\partial \xi_{\bm k\lambda}}\right).
\label{eq:35}
\end{equation}
In this expression, $\tau_{\bm k}$ is the quasiparticle lifetime, 
and $v_{\bm k}$ represents the electron group velocity. The effect of the interaction between conduction electrons and Kondo impurities enters through the quasiparticle lifetime.

We identify the quasiparticle scattering rate as 
\begin{equation}1/\tau_{\bm k\lambda}(\xi_{\bm k\lambda},T)=2\Gamma-2N_{imp}{\rm Im} \mathcal T_{\bm k\lambda}(\xi_{\bm k\lambda}+i\Gamma,T).
\label{eq:36}
\end{equation}

In Eqs.~(\ref{eq:35}) and (\ref{eq:36}), a finite broadening parameter $\Gamma$ is included in the scattering rate. This broadening shifts the poles of the retarded and advanced Green's functions away from the real frequency axis, regularizes the auxiliary Green's functions entering the $T$-matrix, and ensures that the resistivity $\rho^{\gamma\gamma}(T)$ remains finite and well defined. We also explicitly include the total number of impurities  $N_{imp}$. This is justified in the dilute-impurity limit, where correlations between different impurities can be neglected, allowing the impurities to be treated as independent scatterers.

The conductivity expression in Eq.~(\ref{eq:35}) is invariant under the simultaneous interchange of the momentum components $k_x$ and $k_y$ and their corresponding masses $m_x$ and $m_y$. This invariance originates from the symmetry of the energy spectrum $\xi_{m_x,m_y}(k_x,k_y)=\xi_{m_y,m_x}(k_y,k_x)$. Under this exchange symmetry in Eq.~(\ref{eq:35}), one obtains $\sigma^{xx}(m_x,m_y)=\sigma^{yy}(m_y,m_x)$. The same symmetry therefore applies to the resistivities, yielding $\rho^{xx}(m_x,m_y)=\rho^{yy}(m_y,m_x)$. Consequently, varying $m_y$ at fixed $m_x$ in $\rho^{xx}$ produces the same dependence as varying $m_x$ at fixed $m_y$ in $\rho^{yy}$. Conversely, varying $m_x$ at fixed $m_y$ in $\rho^{xx}$ corresponds to varying $m_y$ at fixed $m_x$ in $\rho^{yy}$. In both cases, the mass that is not varied is kept at the same value.

\begin{figure*}[t]
  \subfigure[]  {\includegraphics[width=0.49\linewidth]{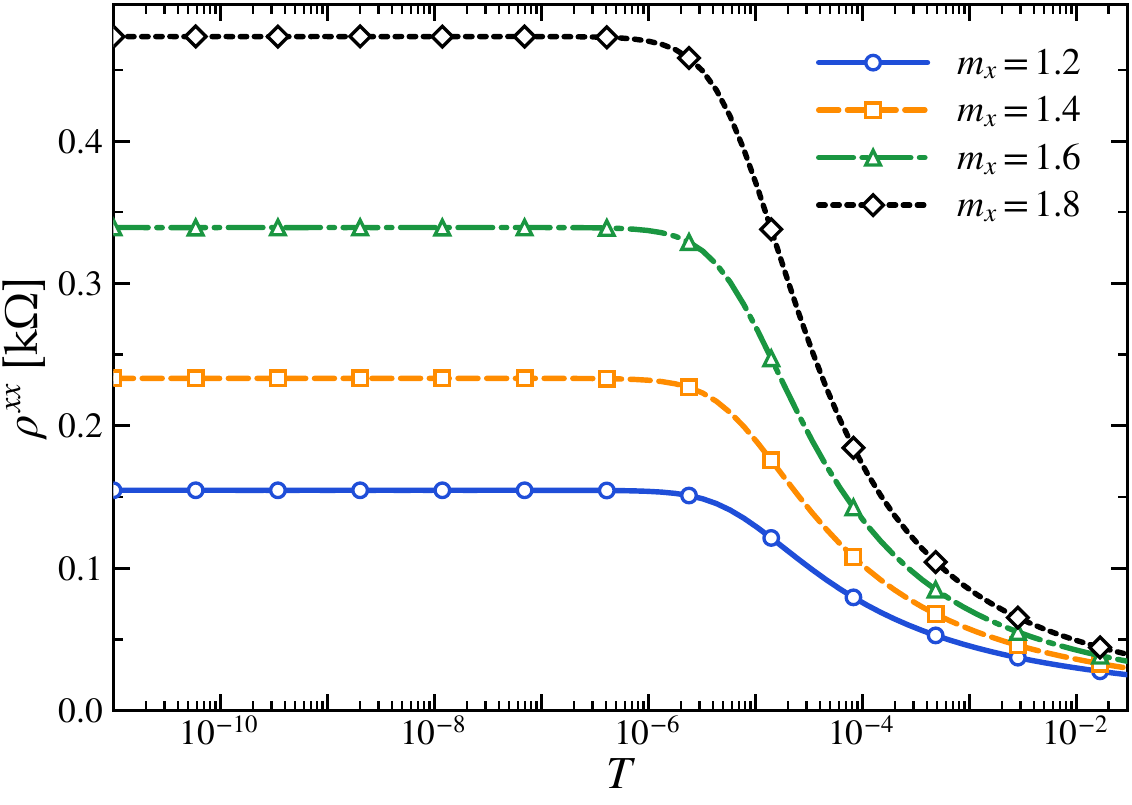}}
  \subfigure[]
{\includegraphics[width=0.49\linewidth]{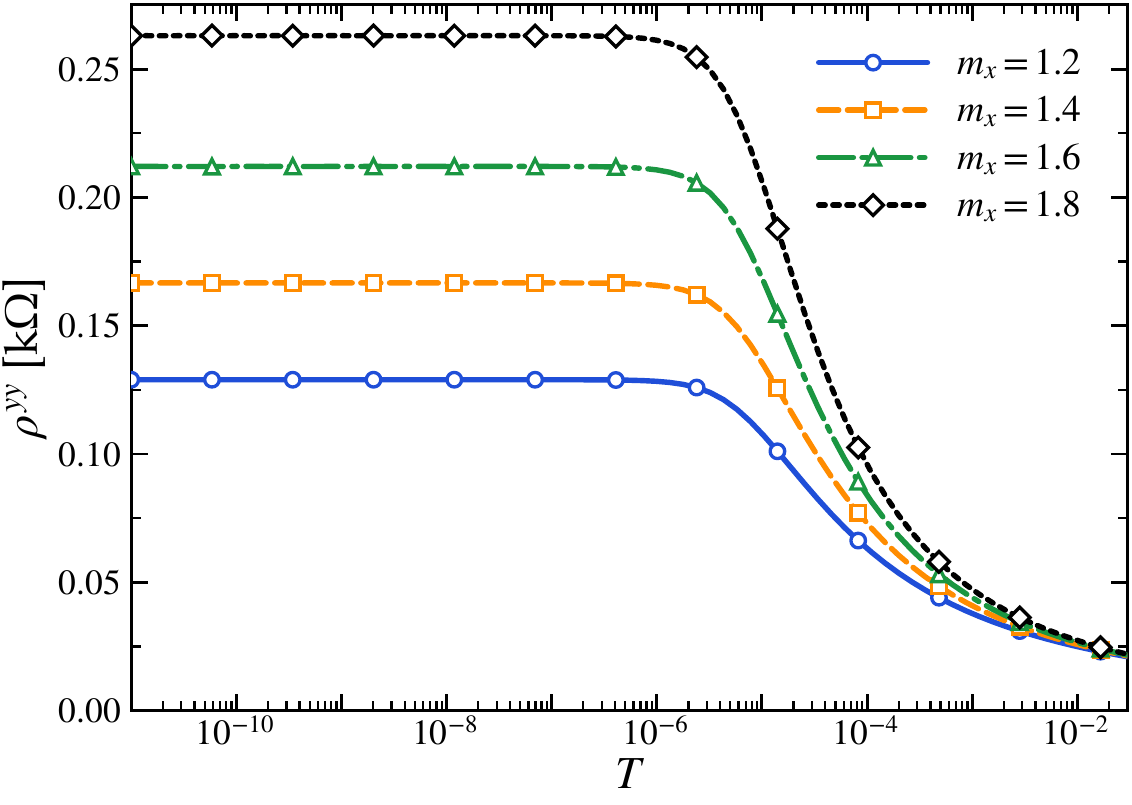}}
\caption{Temperature dependence of the longitudinal resistivity components (a) $\rho^{xx}(T)$ and (b) $\rho^{yy}(T)$ for several values of the effective mass $m_x$, calculated using the Green's function method. The temperature is expressed in dimensionless units as $T\equiv k_BT[{\rm K}]/W_0$, where $W_0=1\;{\rm eV}$. The parameters are $\alpha=10 \,{\rm \mu eV \cdot \AA}^2,\;J_{sd}=-0.4\;{\rm eV},\;\mu=0.25\; {\rm eV}$, $D=0.5\; {\rm eV}$, $\;\Gamma=0.01\;{\rm meV}$, \;$n_{imp}=10^{-2}$, and $\;m_y=1$. The effective masses $m_x$ and $m_y$ are expressed in units of the free-electron mass $m_e$.}
  \label{fig:4}
  \end{figure*}


\subsection{ Green's function method}
\label{sec:5A}
We first evaluate the resistivity within the Green's function formalism. To this end, we introduce the impurity $T$-matrix through Dyson's equation
\begin{equation}\mathcal G_{\bm k \lambda;\bm k'\lambda'}
= g_{\bm k\lambda}\delta_{\bm k\bm k'}\delta_{\lambda \lambda'}+ g_{\bm k\lambda}\mathcal T_{\bm k\lambda;\bm k'\lambda'}g_{\bm k'\lambda'}.
\label{eq:37}
\end{equation}
Here, the dressed Green's function in helicity basis is defined as $\mathcal G_{\bm k\lambda; \bm k'\lambda'}(i\omega_n)=\langle\!\langle \gamma_{\bm k\lambda}|\gamma^\dagger_{\bm k'\lambda'} \rangle\!\rangle$, whereas the free Green's function is given by $g_{\bm k\lambda}(i\omega_n)=(i\omega_n-\xi_{\bm k\lambda})^{-1}$, see Sec.~\ref{sec:4} for details.

Using the transformation from the electron operators $C_{\bm k\sigma}$ to the helicity operators $\gamma_{\bm k\lambda}$ introduced in Sec.~\ref{sec:4}, the sum over $\lambda'$ can be written as 
\begin{equation}\sum_{\lambda'}\mathcal G_{\bm k\lambda;\bm k'\lambda'}(i\omega_n)=\mathcal G_{\bm k\bm k' }(i\omega_n)+\lambda e^{i2 \varphi_{\bm k}}\mathcal F_{\bm k\bm k'}(i\omega_n).
\label{eq:38}
\end{equation}
Summing Eq.~(\ref{eq:37}) over $\lambda'$, we obtain
\begin{equation}\sum_{\lambda'}\mathcal G_{\bm k\lambda;\bm k'\lambda'}=g_{\bm k\lambda}\delta_{\bm k\bm k'}+g_{\bm k\lambda}\sum_{\lambda'}\mathcal T_{\bm k\lambda;\bm k'\lambda'}g_{\bm k'\lambda'}.
\label{eq:39}
\end{equation}
In extracting the $T$-matrix $\mathcal T_{\bm k\lambda;\bm k'\lambda'}$, we first substitute Eqs.~(\ref{eq:14a}) and (\ref{eq:14b}) into Eq.~(\ref{eq:38}), after which we equate the resulting expression with Eq.~(\ref{eq:39}) and use
\begin{equation} g_{\bm k'}=\frac{1}{2}\sum_{\lambda'=\pm}g_{\bm k'\lambda'}, \qquad h_{\bm k'\pm}=\frac{e^{\pm 2i\varphi_{\bm k'}}}{2}\sum_{\lambda'=\pm}\lambda'g_{\bm k'\lambda'}.
\label{eq:40}
\end{equation}
This procedure yields the following expression for the $T$-matrix element
\vspace{-10mm}
\begin{widetext}
$$\mathcal T_{\bm k\lambda;\bm k'\lambda'}(i\omega_n,T)=\frac{J_{sd}}{4N_s}\frac{1}{\mathcal E(i\omega_n)}\Bigg[\frac{ \Xi(i\omega_n)[1-\mathcal B(i\omega_n)]+\Phi(i\omega_n) \mathcal I(i\omega_n)+\lambda \lambda'e^{2i(\varphi_{\bm k}-\varphi_{\bm k'})}[ \Pi(i\omega_n)\mathcal R(i\omega_n)+\Omega(i\omega_n)(1-\mathcal U(i\omega_n))]}{(1-\mathcal U(i\omega_n))(1-\mathcal B(i\omega_n))-\mathcal R(i\omega_n)\mathcal I(i\omega_n)}$$
\vspace{-3mm}
\begin{equation}+\frac{\lambda e^{2i\varphi_{\bm k}}\left[ \Pi(i\omega_n)(1-\mathcal B(i\omega_n))+\Omega(i\omega_n)\mathcal I(i\omega_n)\right]+\lambda' e^{-2i\varphi_{\bm k'}}\left[\Xi(i\omega_n)\mathcal R(i\omega_n)+\Phi(i\omega_n)(1-\mathcal U(i\omega_n))\right]}{(1-\mathcal U(i\omega_n))(1-\mathcal B(i\omega_n))-\mathcal R(i\omega_n)\mathcal I(i\omega_n)}\Bigg],
\label{eq:41}
\end{equation}
\end{widetext}
where the explicit forms of the auxiliary functions $\Xi(i\omega_n), \Pi(i\omega_n), \Phi(i\omega_n)$, and $\Omega(i\omega_n)$ are given in Eqs.~(\ref{eq:C9a})--(\ref{eq:C9d}).

Having obtained an explicit expression for $\mathcal T_{\bm k\lambda;\bm k'\lambda'}$, we next determine self-consistently the correlators $n_{\bm k}$ and $m_{\bm k}$ entering the auxiliary functions. They are defined as follows (see Appendix~\ref{appendix:B} for details)
\begin{equation}
n_{\bm k}=\sum_{\bm p}\langle C^\dagger_{\bm p\uparrow}C_{\bm k\uparrow}
\rangle,\qquad m_{\bm k}/3=\sum_{\bm p}\langle C^\dagger_{\bm p\uparrow}C_{\bm k\downarrow}S^-\rangle.
\label{eq:42}
\end{equation}
The correlators $n_{\bm k}$ and $m_{\bm k}$ are obtained from the single-particle Green's function $\mathcal G_{\bm k\bm p}$ and the spin-flip Green's function $\langle\!\langle C_{\bm k\downarrow}S^-|C^\dagger_{\bm p\uparrow}\rangle\!\rangle$, respectively. The latter is evaluated by substituting Eqs.~(\ref{eq:D12})--(\ref{eq:D14}) into Eq.~(\ref{eq:B3}). Applying the spectral theorem to the corresponding Green's functions, we obtain
\begin{equation}n_{\bm k}=-\frac{1}{\pi}\int d\omega n_F(\omega){\rm Im}\sum_{\bm p}\mathcal G_{\bm k\bm p}(\omega+i\Gamma),
\label{eq:43}
\end{equation}
and
\begin{equation}m_{\bm k}=-\frac{3}{\pi}\int d\omega n_F(\omega){\rm Im}\sum_{\bm p}\langle\!\langle C_{\bm k\downarrow}S^-|C^\dagger_{\bm p\uparrow} \rangle\!\rangle_{\omega+i\Gamma}.
\label{eq:44}
\end{equation}

\begin{figure*}[t]
  \subfigure[]
  {\includegraphics[width=0.49\linewidth]{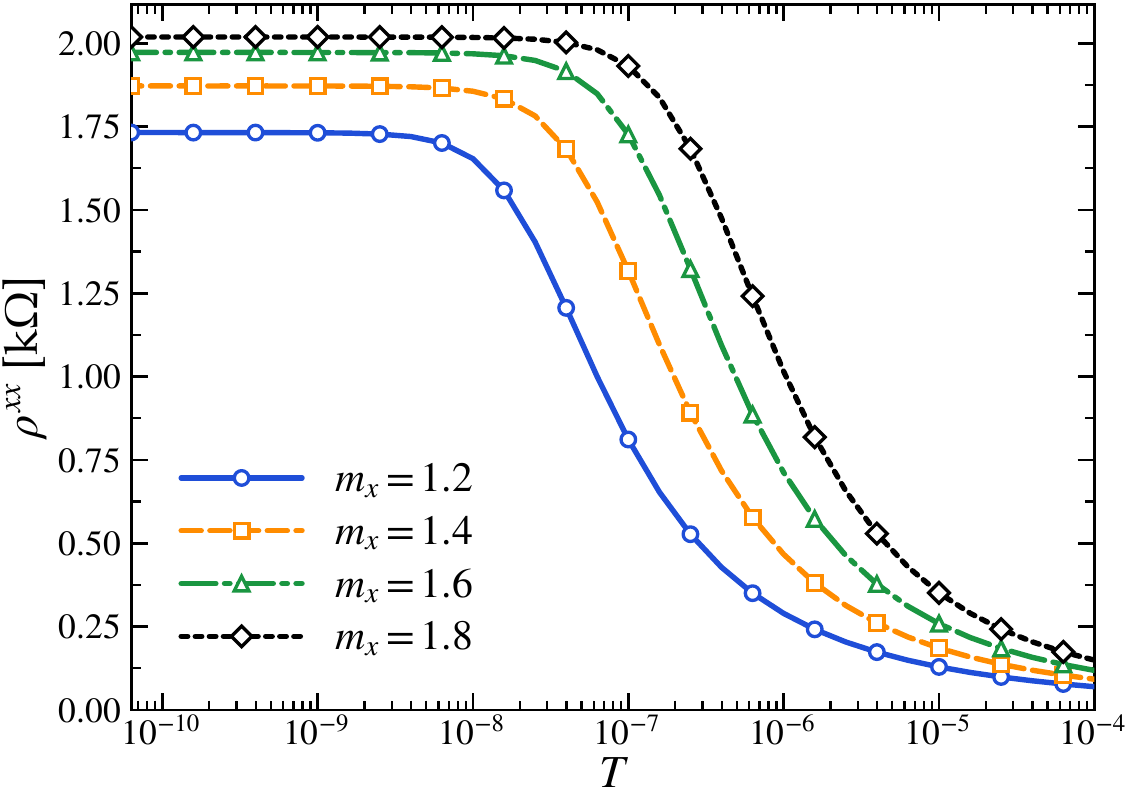}}
  \subfigure[]{\includegraphics[width=0.49\linewidth]{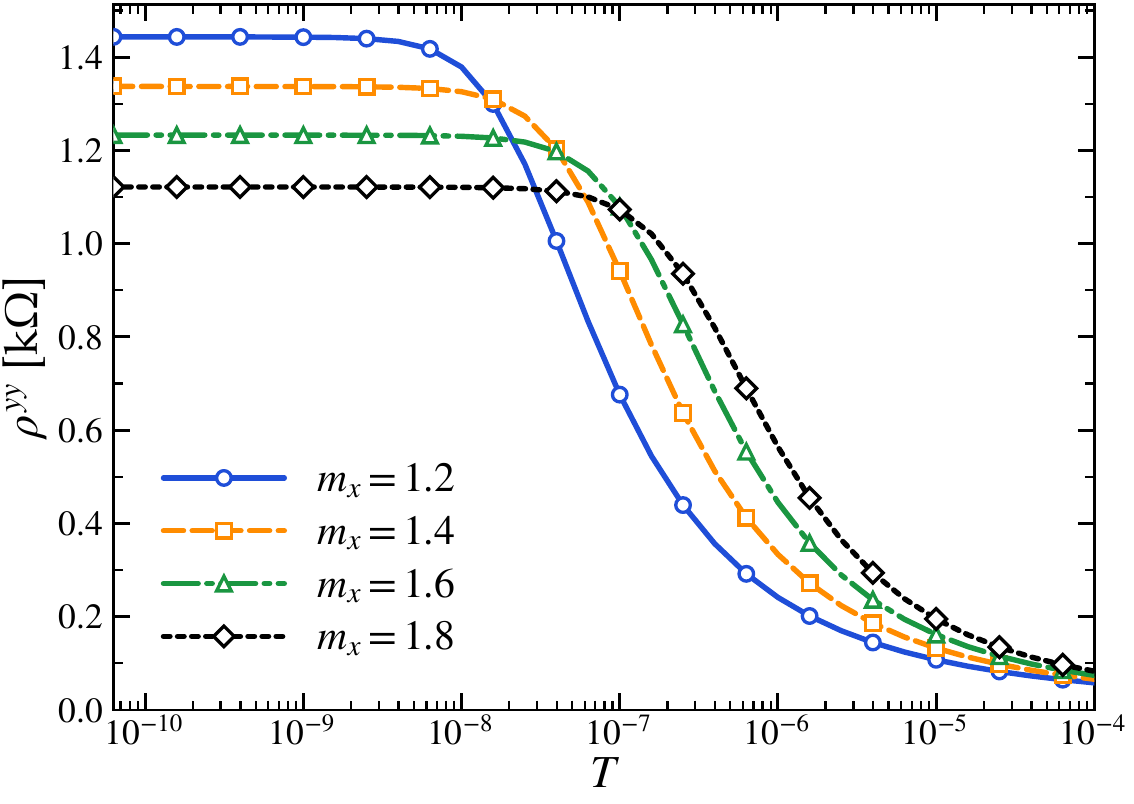}}
 \caption{Temperature dependence of the longitudinal resistivity components (a) $\rho^{xx}(T)$ and (b) $\rho^{yy}(T)$ for several values of the effective mass $m_x$, calculated using the Ljubljana NRG framework \cite{zitko2009,zitko2009_1,bulla2008}. The temperature is expressed in dimensionless units as $T\equiv k_BT[{\rm K}]/W_0$, where $W_0=1\;{\rm eV}$. The parameters are $\alpha=10 \,{\rm  \mu eV \cdot \AA}^2,\;J_{sd}=0.4\;{\rm eV},\;\mu=0.25\; {\rm eV}$, $D=0.5\; {\rm eV}$, $\;\Gamma=0.01\;{\rm meV}$, \;$n_{imp}=10^{-2}$, and $\;m_y=1$. The effective masses $m_x$ and $m_y$ are expressed in units of the free-electron mass $m_e$.}
  \label{fig:5}
  \end{figure*}

The coupled self-consistency equations for $n_{\bm k}$ and $m_{\bm k}$ were solved by fixed-point iterations. We decomposed these quantities as $n_{\bm k}=n_{\bm k}^{(0)}+\delta n_{\bm k}$ and $m_{\bm k}=m_{\bm k}^{(0)}+\delta m_{\bm k}$, where $n_{\bm k}^{(0)}$ is the undressed occupation given by Eq.~(\ref{eq:18}) and $m_{\bm k}^{(0)}=0$. The latter follows because the mixed electron-impurity spin-flip correlation vanishes in the absence of exchange coupling $J_{sd}$. Interactions therefore generate $\delta n_{\bm k}$ and $\delta m_{\bm k}$. At each iteration, these corrections were recalculated from the dressed single-particle Green's function $\mathcal G_{\bm k\bm p}$ and the spin-flip Green's function $\langle \!\langle C_{\bm k\downarrow}S^-|C^\dagger_{\bm p\uparrow}\rangle\!\rangle$, respectively. Convergence was declared when the maximum absolute differences between the recalculated and input values over all $\bm k$ were smaller than $10^{-5}$ for both $\delta n_{\bm k}$ and $\delta m_{\bm k}$.

In evaluating the resistivity, we use the diagonal $T$-matrix element $\mathcal T_{\bm k\lambda}\equiv \mathcal T_{\bm k\lambda;\bm k\lambda}$ obtained from Eq.~(\ref{eq:41}). We introduce the reduced impurity $T$-matrix ${\widetilde {\mathcal T}_{\bm k\lambda}}=N_s \mathcal T_{\bm k\lambda}$ together with the impurity concentration $n_{imp}=N_{imp}/N_s$ and substitute these definitions into Eq.~(\ref{eq:36}). We then define the rescaled momentum variables $K_\gamma=k_\gamma/\sqrt{m_\gamma}$, with $\gamma=x, y$, which satisfy $K^2=2(\xi+\mu)$, and rewrite the momentum summation as \cite{coleman2015,yanagisawa2005}
\begin{equation}
\frac{1}{N_s}\sum_{\bm k}(\hdots)\to \nu_{2D}\int_{0}^{2\pi}\frac{d\varphi}{2\pi}\int_{-\mu}^{D-\mu}d\xi(\hdots).
\label{eq:45}
\end{equation}
We further use the helicity energy $\xi_{\bm k\lambda}$ as the integration variable. We also note that the angular variable $\varphi$ introduced in Eq.~(\ref{eq:45}) differs from the momentum angle $\varphi_{\bm k}$ appearing in Eq.~(\ref{eq:41}). These angles are defined by $\varphi_{\bm k}={\rm atan}2\,(k_y,k_x)$ and $\varphi={\rm atan}2\,(K_y,K_x)$  and are related through $\tan \varphi_{\bm k}=\sqrt{m_y/m_x}\tan \varphi$. Finally, we obtain the following expression for the resistivity a
\begin{widetext}
\begin{equation}\rho^{\gamma\gamma}(T)=\frac{2m_\gamma}{e^2\nu_{2D}}\left[\sum_{\lambda=\pm}\int_{0}^{2\pi}\frac{d\varphi}{2\pi}M^\gamma_\varphi \left(\frac{1+2\lambda \alpha m_\gamma}{1+\lambda \alpha L_\varphi}\right)^2 \int_{-\mu/T}^{[D(1+\lambda \alpha L_\varphi)-\mu]/T}dx (\mu+xT)\frac{{\rm sech}^2(x/2)}{2\Gamma-2n_{imp} {\rm Im}{\widetilde {\mathcal T}}_{\bm k\lambda}(xT+i\Gamma, T)}\right]^{-1}.
    \label{eq:46}
    \end{equation}
\end{widetext}
In Eq.~(\ref{eq:46}), we introduced the angular functions $M^x_\varphi=\cos^2\varphi $, $M^y_\varphi=\sin^2\varphi$, and $L_\varphi=2(m_x\cos^2\varphi+m_y\sin^2\varphi)$.

We now analyze the temperature dependence of the resistivity shown in Fig.~\ref{fig:4}. In the low-temperature regime $T\ll \Gamma$, the resistivity is governed by the quasiparticle broadening $\Gamma$ and the quadratic spin splitting parameter $\alpha$, leading to the behavior $\rho^{\gamma\gamma}= \rho^{\gamma\gamma}_0(\alpha)-b(\alpha)T^2$. Here, $\Gamma$ determines the quadratic temperature dependence, while $\alpha$ quantitatively modifies the resistivity through the helicity splitting of the conduction band. Thus, $\alpha$ modifies the impurity contribution to the resistivity through its interplay with $J_{sd}$ and $\Gamma$, rather than acting as an independent scattering mechanism. At $T\sim \Gamma$, the resistivity crosses over from the low-temperature regime to the high-temperature Kondo regime. For $T\gg \Gamma$, the impurity spin-flip contribution exhibits the characteristic logarithmic temperature dependence $\rho^{\gamma\gamma}\propto -\ln T$.

The resistivity is also proportional to the impurity concentration, indicating that a higher concentration of Kondo impurities enhances the resistivity. This is consistent with the increase in the scattering rate $1/\tau_{\bm k}$, since a larger number of impurities provides more scattering centers for conduction electrons.


\subsection{ NRG calculation}
In this section, we discuss a more accurate method for calculating the resistivity of dilute Kondo alloys with quadratic spin splitting. This approach provides a quantitatively reliable description of the resistivity  across the entire temperature range, from the high-temperature regime, through the crossover at $T\sim T_K$, down to the low-temperature strong-coupling regime, $T\ll T_K$. We use the NRG parameters specified in Sec.~\ref{sec:4B}.

We begin by expressing the quasiparticle scattering rate in terms of the impurity $T$-matrix as follows
\begin{equation}
\frac{1}{\tau_{\bm k\lambda}(\omega,T)}\Big|_{\omega=\xi_{\bm k\lambda}}=2\Gamma-2 n_{imp}{\rm Im}\left[ \langle u_{\bm k\lambda}|{{\bm{\mathcal T}}}^R(\omega,T)|u_{\bm k\lambda}\rangle\right]\Big|_{\omega=\xi_{\bm k\lambda}}.
\label{eq:47}
\end{equation}
To evaluate the $T$-matrix element in Eq.~(\ref{eq:47}), we use the spin-basis states and helicity eigenstates given by
\begin{equation}|\chi_{s}\rangle=\frac{1}{\sqrt{2}}\begin{pmatrix}
    1\\\\
    s
\end{pmatrix},\qquad |u_{\bm k\lambda}\rangle=\frac{1}{\sqrt{2}}\begin{pmatrix}
    1\\\\
    \lambda e^{-2i\varphi}
\end{pmatrix},
\label{eq:48}
\end{equation}
where $s=\pm 1$.

By projecting the helicity eigenstate $|u_{\bm k\lambda}\rangle$ onto the fixed spin basis $|\chi_{\pm}\rangle$, in which the impurity $T$-matrix is diagonal, we obtain the corresponding $T$-matrix element in the helicity basis as 
$$\langle u_{\bm k\lambda}|{\bm{\mathcal T}}^R(\omega,T)|u_{\bm k\lambda}\rangle$$
\begin{equation}=|\langle\chi_+|u_{\bm k\lambda}\rangle|^2  \mathcal T^R_+(\omega,T)+|\langle\chi_-|u_{\bm k\lambda}\rangle|^2  \mathcal T^R_-(\omega,T).
\label{eq:49}
\end{equation}
Here, $\mathcal T^R_\pm(\omega,T)=\langle \chi_\pm| {{\bm{\mathcal T}}}^R(\omega,T)|\chi_\pm\rangle$ denote the diagonal spin-resolved components of the retarded impurity $T$-matrix in the $\chi_\pm$ basis. The corresponding spectral densities are calculated within the Ljubljana NRG framework \cite{zitko2009,zitko2009_1},  using the $z$-averaging procedure described in Sec.~\ref{sec:4B}, and are defined by
\begin{equation} T_\pm(\omega,T)=-\frac{1}{\pi}{\rm Im} \mathcal T^R_\pm(\omega,T).
\label{eq:50}
\end{equation}

In the NRG calculation, frequency and temperature are expressed through the dimensionless variables $\widetilde{\omega}=\omega/W$ and $T^{NRG}=k_BT/W$, respectively. The physical spectral density entering the scattering rate is then obtained as 
\begin{equation}
 T_\pm(\omega,T)=\frac{W}{N} T^{NRG}_\pm\left(\frac{\omega}{W},\frac{k_B T}{W}\right),
\label{eq:51}
\end{equation}
where $N=D\nu_{2D}.$

Substituting Eq.~(\ref{eq:49}) into Eq.~(\ref{eq:47}), we obtain the scattering rate
{\small
\begin{equation*}\frac{1}{\tau_{\bm k\lambda}(\omega,T)}=2\Gamma+2\pi n_{imp}\frac{W}{N}\Bigg[ \frac{1+\lambda \cos (2\varphi_{\bm k})}{2} T^{NRG}_+\left(\frac{\omega}{W},\frac{k_B T}{W}\right)
\end{equation*}
\begin{equation}+\frac{1-\lambda \cos(2\varphi_{\bm k})}{2} T^{NRG}_-\left(\frac{\omega}{W},\frac{k_B T}{W}\right)\Bigg].
\label{eq:52}
\end{equation}
}

Finally, inserting  Eq.~(\ref{eq:52}) into Eq.~(\ref{eq:46}) and noting that the angular variables $\varphi_{\bm k}$ appearing in Eq.~(\ref{eq:52}) and $\varphi$ used in Eq.~(\ref{eq:46}) are different, see Sec.~\ref{sec:5A} for details, we obtain the temperature dependence of the resistivity $\rho^{\gamma\gamma}(T)$ within the NRG framework, as shown in Fig.~\ref{fig:5}.

The resistivity curves in Fig.~\ref{fig:5} exhibit a smooth crossover from the low-temperature strong-coupling regime to the high-temperature Kondo regime. At the lowest temperatures $T\ll T_K$, Kondo correlations strongly screen the impurity spins, and the resistivity exhibits the quadratic temperature dependence $\rho^{\gamma\gamma}(T)=\rho_0^{\gamma\gamma}(\alpha)-b(\alpha)T^2$. With increasing temperature toward $T_K$, Kondo screening becomes weaker, and in the high-temperature regime  $T\gg T_K$, the impurity spins remain essentially unscreened, while the resistivity exhibits the characteristic Kondo temperature dependence.

\begin{figure}[t!]
  \subfigure[]
  {\includegraphics[width=1\linewidth]{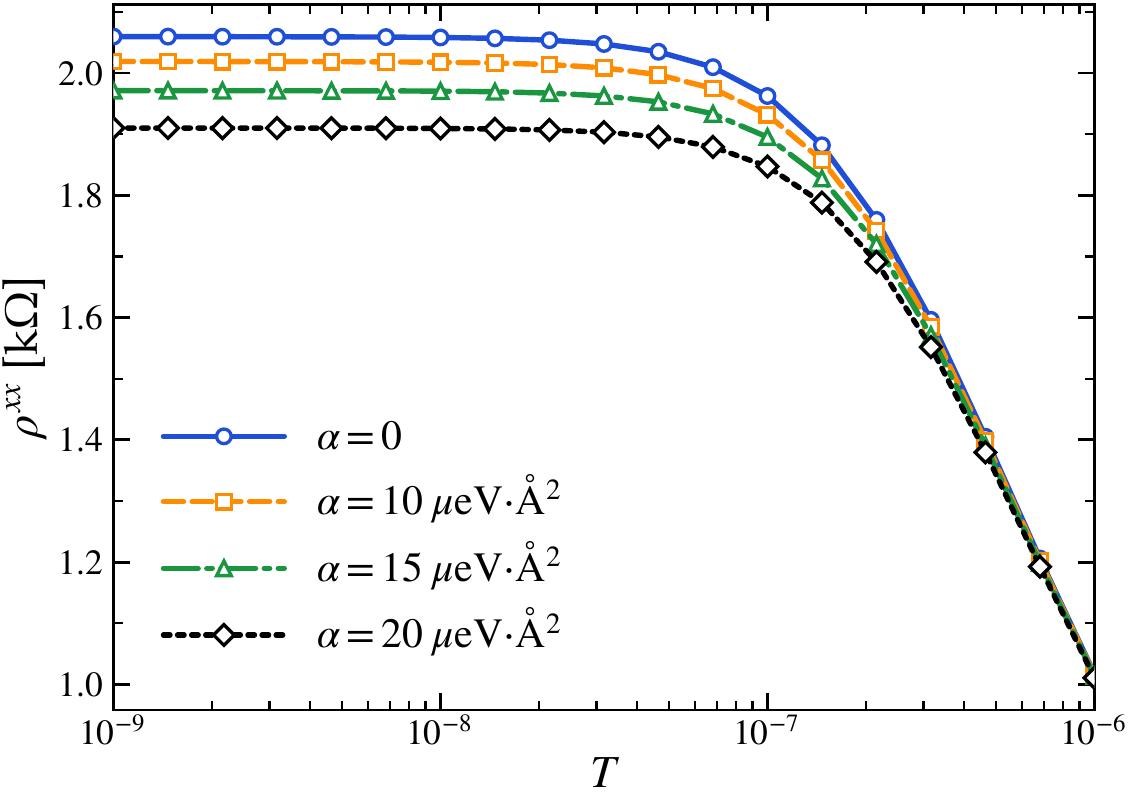}}
  \subfigure[] 
   {\includegraphics[width=1\linewidth]{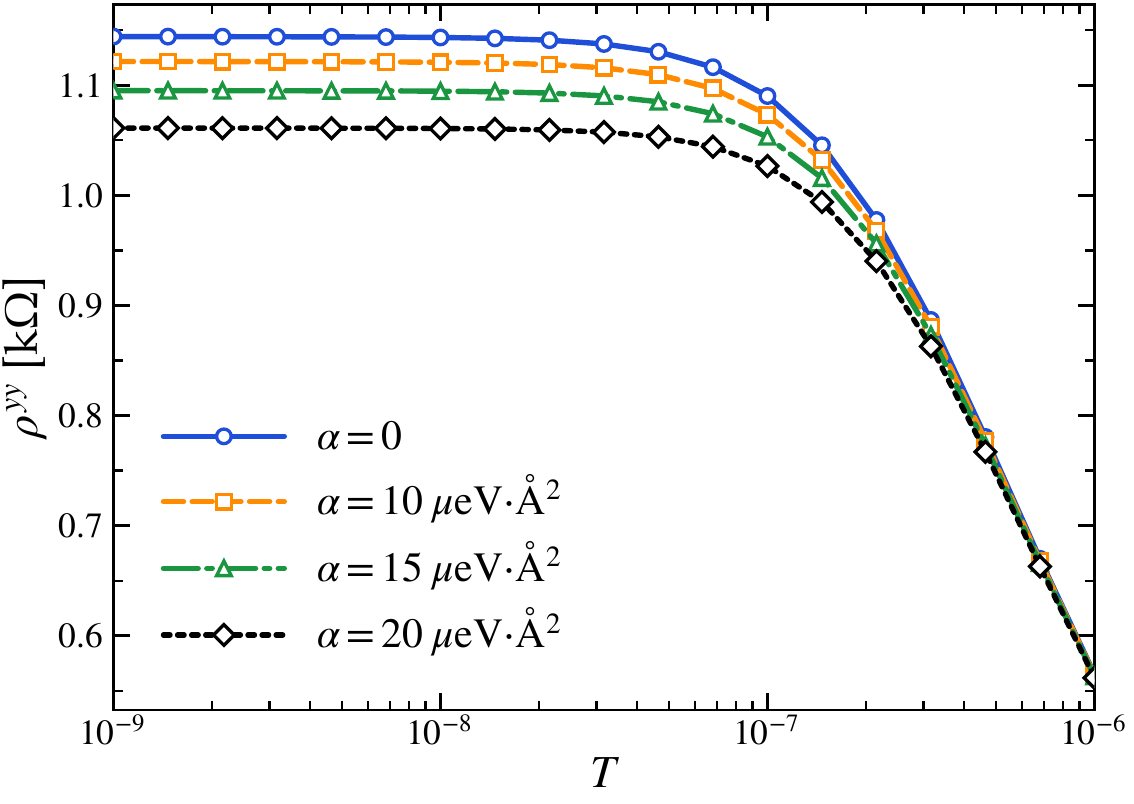}}
 \caption{Temperature dependence of the longitudinal resistivity components (a) $\rho^{xx}(T)$ and (b) $\rho^{yy}(T)$ for several values of the quadratic spin-splitting coupling $\alpha$, calculated using the Ljubljana NRG framework \cite{zitko2009,zitko2009_1,bulla2008}. The temperature is expressed in dimensionless units as $T\equiv k_BT[{\rm K}]/W_0$, where $W_0=1\;{\rm eV}$. The parameters are $J_{sd}=0.4\;{\rm eV},\;\mu=0.25\; {\rm eV}$, $D=0.5\; {\rm eV}$, $\;\Gamma=0.01\;{\rm meV}$, \;$n_{imp}=10^{-2}$, $m_x=1.8$, and $\;m_y=1$. The effective masses $m_x$ and $m_y$ are expressed in units of the free-electron mass $m_e$.}
  \label{fig:6}
  \end{figure}

The influence of quadratic spin splitting on the resistivity depends crucially on mass anisotropy. As shown in Fig.~\ref{fig:6}, varying $\alpha$ in the anisotropic system changes the magnitude of the low-temperature resistivity without altering its quadratic temperature dependence, whereas the resistivity remains unchanged upon varying $\alpha$ in the isotropic system. This $\alpha$ dependence reflects the modification of Kondo screening by quadratic spin splitting. Within the strong-coupling regime, stronger Kondo screening produces a larger impurity contribution to the resistivity. Increasing $\alpha$ weakens the screened state and therefore reduces the low-temperature resistivity, yielding $\rho^{\gamma\gamma}(\alpha=0)>\rho^{\gamma\gamma}(\alpha\neq 0)$. At higher temperatures, the resistivity becomes less sensitive to $\alpha$, since Kondo screening is already weakened and its additional suppression by quadratic spin splitting has little effect in the regime $T\gg T_K$.

\section{Discussion}
\label{sec:6}
The suppression of Kondo screening with increasing $\alpha$ is reflected in the normalized Kondo temperature $T_K(\alpha)/T_K^{(0)}$ and the impurity screening function $p(\alpha)$, both of which  decrease monotonically, as shown in Figs.~\ref{fig:2} and \ref{fig:3}. While $T_K(\alpha)/T_K^0$ measures the reduction of the characteristic screening scale, the function $p(\alpha)$ reflects the growth of the impurity polarization along the $x$-axis associated with weakened Kondo screening. Microscopically, this behavior is governed by two $\alpha$-dependent energy scales, the separation $\Delta_{split}$ between the helicity branches and the effective field $h_{eff}$, generated in the presence of $s$-$d$ exchange interaction and mass anisotropy. The helicity splitting $\Delta_{split}$ weakens the inter-helicity spin-flip scattering, thereby lowering $T_K(\alpha)$ and reducing the ability of conduction electrons to screen the impurity moment, whereas $h_{eff}$ favors a finite impurity polarization. Their influence becomes pronounced in the corresponding crossover regimes $\Delta_{split}\sim k_BT_K$ and $ |h_{eff}|\sim k_BT_K$. The microscopic picture shows how quadratic spin splitting, in the presence of mass anisotropy, weakens Kondo screening, thereby enhancing the impurity polarization.

\begin{figure}[t]
{\includegraphics[width=1\linewidth]{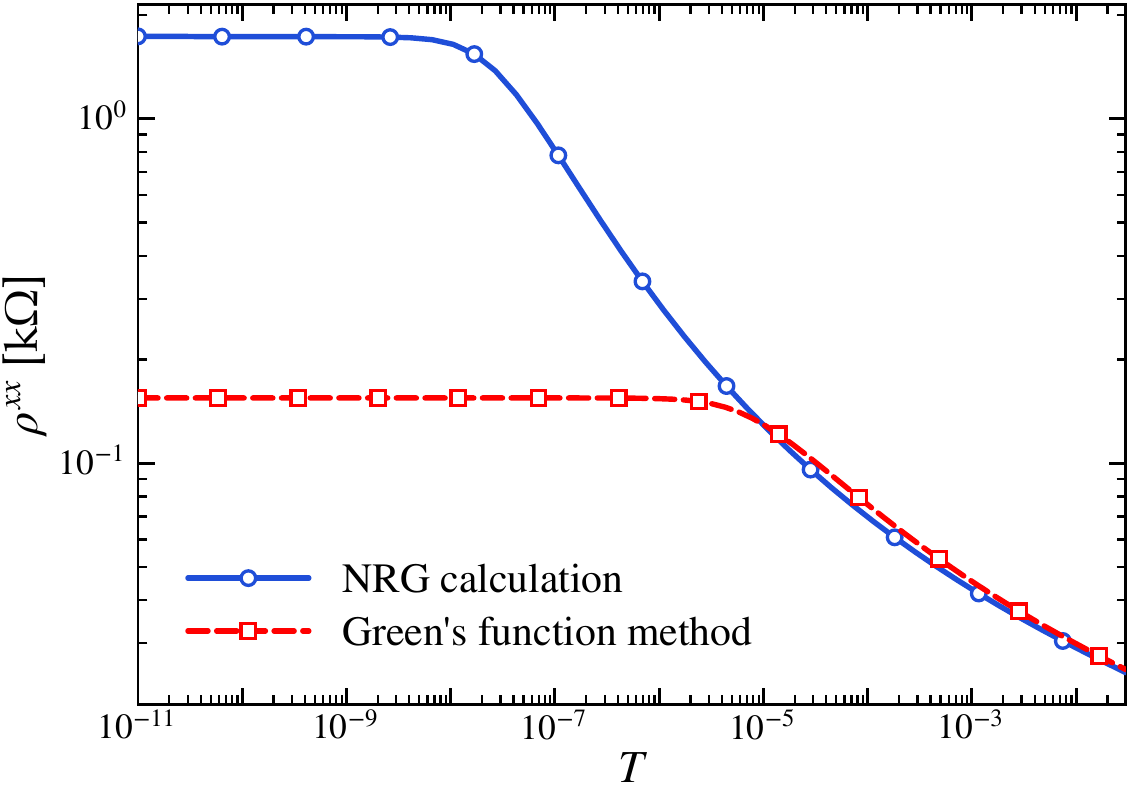}}
\caption{Comparison of the resistivity obtained using the Green's function method and NRG calculation. The temperature is expressed in dimensionless units as $T\equiv k_BT[{\rm K}]/W_0$, where $W_0=1\;{\rm eV}$. The parameters are the same as those used in Figs.~\ref{fig:4} and \ref{fig:5}, with $m_x=1.2$ and $m_y=1$. The effective masses $m_x$ and $m_y$ are expressed in units of the free-electron mass $m_e$.} 
\label{fig:7}
  \end{figure}

The resistivity calculated using the self-consistent Green's function method is shown in Fig.~\ref{fig:4}. Although this method provides a reliable description in the high-temperature regime $T\gg T_K$ and qualitatively reproduces the reduction of the resistivity with increasing spin-splitting coupling $\alpha$, it does not provide a quantitatively accurate description in the low-temperature regime $T\ll T_K$ or near the Kondo crossover $T\sim T_K$. Further improvement of the analytical framework would require extending the decoupling scheme presented in Eqs.~(\ref{eq:B13})--(\ref{eq:B22}) to retain higher-order correlation functions, which could provide a more accurate description of the many-body correlations involved in the development of the strong-coupling state. By contrast, the NRG calculations shown in Fig.~\ref{fig:5} provide a reliable description of the resistivity over the full temperature range. A direct comparison between the results obtained using the Green's function and NRG approaches is presented in Fig.~\ref{fig:7}.

The present low-energy effective model, see Hamiltonian~(\ref{eq:2}), is motivated by BaSbPt \cite{liu2024}, a hexagonal compound whose symmetry permits a quadratic momentum-dependent spin texture, making it a natural candidate for experimental investigation of the Kondo transport discussed here. Similar quadratic textures have also been reported in MnTe$_2$ \cite{zhu2024} and AgTlP$_2$Se$_6$ \cite{zhou2026}. Although the  effective models for these materials differ from the present one, they belong to the same general class of systems exhibiting quadratic spin splitting. Our results should therefore remain qualitatively relevant for such materials, while the quantitative behavior depends on the specific electronic spectrum, density of states, and dimensionality.

While this study is restricted to the simplest case of conduction fermions with spin $s=1/2$, the present formalism can be generalized to systems with higher spins \cite{xu2022}. Such an extension would enable a systematic investigation of how larger spin representations and additional spin-scattering channels modify the temperature dependence of the resistivity and the associated temperature scales. Understanding transport in such systems would be valuable for the theoretical description of a broader class of Kondo materials and for identifying promising candidates in which multiple spin-scattering channels play an important role in determining the transport properties. A detailed investigation of these systems will be reported elsewhere.

The present results provide a theoretical basis for identifying Kondo transport signatures in anisotropic materials with quadratic spin-splitting textures.  Although direct experimental studies of quadratic spin-splitting coupling remain limited, the magnitude of linear Rashba spin-orbit coupling is well known to be tunable by gate voltage \cite{nitta1997,takase2017}, charge doping \cite{chen2020}, and strain \cite{xiang2019,bhumla2021}.  By analogy, these external parameters may offer routes to tuning quadratic spin splitting and hence the Kondo resistivity in materials exhibiting quadratic spin textures, as well as in $d$-wave altermagnets \cite{golub2025,herasymchuk2025}, which exhibit quadratic spin splitting with a different momentum-space structure.

\section{Conclusion}
\label{sec:7}
In this paper, we studied the Kondo effect in a two-dimensional anisotropic electronic system with a quadratic spin texture and examined its influence on charge transport. Using the Green's function method and NRG calculation, we evaluated the temperature-dependent resistivity components along the two inequivalent in-plane directions,  $\rho^{xx}(T)$ and $\rho^{yy}(T)$, as shown in Figs.~\ref{fig:4} and \ref{fig:5}. We also found that both the Kondo scale $T_K(\alpha)$ and the impurity screening function $p(\alpha)$ are strongly suppressed with increasing quadratic spin-splitting coupling $\alpha$, as shown in Figs.~\ref{fig:2} and \ref{fig:3}.

\onecolumngrid
\appendix

\section{Equations of motion for the Green's functions}
\label{appendix:A}

We first collect the operator identities for the localized spin-$1/2$ operators:
\begin{equation}
S^\pm S^z=\mp\frac{1}{2}S^\pm,\qquad S^zS^\pm=\pm \frac{1}{2}S^\pm,\qquad 
S^+S^-=\frac{3}{4}+S^z-(S^z)^2,\qquad S^-S^+=\frac{3}{4}-S^z-(S^z)^2,
\label{eq:A1}
\end{equation}
and
\begin{equation}
[S^\pm,S^z]=\mp S^\pm,\qquad [S^z,S^\pm]=\pm S^\pm,\qquad [S^+,S^-]=2S^z.
\label{eq:A2}
\end{equation}
For the fermionic operators, we use the anticommutation relation
\begin{equation}
\{C_{\bm k\sigma},C_{\bm k'\sigma'}^\dagger\}=\delta_{\bm k\bm k'}\delta_{\sigma \sigma'},
\label{eq:A3}
\end{equation}
and
\begin{equation}
[AB,C]=A\{B,C\}-\{A,C\}B,
\label{eq:A4}
\end{equation}
Applying the EOM and the commutators given above, we derive the following equations for the Green's functions
{\small
$$(i\omega_n-\xi_{\bm k})\langle\!\langle C_{\bm k\uparrow}S^z|C^\dagger_{\bm k'\uparrow}\rangle\!\rangle=\alpha k_+^2\langle\!\langle C_{\bm k\downarrow}S^z|C^\dagger_{\bm k'\uparrow}\rangle\!\rangle-\frac{J_{sd}}{2 N_s}\sum_{\bm q\bm p}\Big[\langle\!\langle C_{\bm k\uparrow}C^\dagger_{\bm p\downarrow}C_{\bm q\uparrow}S^+|C^\dagger_{\bm k'\uparrow}\rangle\!\rangle-\langle\!\langle C_{\bm k\uparrow}C^\dagger_{\bm p\uparrow}C_{\bm q\downarrow}S^-|C^\dagger_{\bm k'\uparrow}\rangle\!\rangle\Big]$$
\begin{equation}
-\frac{J_{sd}}{2 N_s}\sum_{\bm q}\Big[\langle\!\langle C_{\bm q\uparrow}(S^z)^2|C^\dagger_{\bm k'\uparrow}\rangle\!\rangle +\frac{1}{2}\langle\!\langle C_{\bm q\downarrow}S^-|C^\dagger_{\bm k'\uparrow}\rangle\!\rangle\Big],	
\label{eq:A5}
\end{equation}

$$(i\omega_n-\xi_{\bm k})\langle\!\langle C_{\bm k\downarrow}S^-|C^\dagger_{\bm k'\uparrow}\rangle\!\rangle=\alpha k_-^2\langle\!\langle C_{\bm k\uparrow}S^-|C^\dagger_{\bm k'\uparrow}\rangle\!\rangle-\frac{J_{sd}}{2 N_s}\sum_{\bm q\bm p}\Big[\langle\!\langle C_{\bm k\downarrow}C^\dagger_{\bm p\uparrow}C_{\bm q\uparrow}S^-|C^\dagger_{\bm k'\uparrow}\rangle\!\rangle-\langle\!\langle C_{\bm k\downarrow}C^\dagger_{\bm p\downarrow}C_{\bm q\downarrow}S^-|C^\dagger_{\bm k'\uparrow}\rangle\!\rangle-2\langle\!\langle C_{\bm k\downarrow}C^\dagger_{\bm p\downarrow}C_{\bm q\uparrow}S^z|C^\dagger_{\bm k'\uparrow}\rangle\!\rangle\Big]$$
\begin{equation}-\frac{J_{sd}}{2 N_s} \sum_{\bm q}\Big[\frac{3}{4}\langle\!\langle C_{\bm q\uparrow}|C^\dagger_{\bm k'\uparrow}\rangle\!\rangle-\langle\!\langle C_{\bm q\uparrow}(S^z)^2|C^\dagger_{\bm k'\uparrow}\rangle\!\rangle+\frac{1}{2}\langle\!\langle C_{\bm q\downarrow}S^-|C^\dagger_{\bm k'\uparrow}\rangle\!\rangle+\langle\!\langle C_{\bm q\uparrow}S^z|C^\dagger_{\bm k'\uparrow}\rangle\!\rangle\Big],
\label{eq:A6}
\end{equation}

$$(i\omega_n-\xi_{\bm k})\langle\!\langle C_{\bm k\uparrow}S^-|C^\dagger_{\bm k'\uparrow}\rangle\!\rangle=\alpha k_+^2\langle\!\langle C_{\bm k\downarrow}S^-|C^\dagger_{\bm k'\uparrow}\rangle\!\rangle-\frac{J_{sd}}{2 N_s}\sum_{\bm q\bm p}\Big[\langle\!\langle C_{\bm k\uparrow}C^\dagger_{\bm p\uparrow}C_{\bm q\uparrow}S^-|C^\dagger_{\bm k'\uparrow}\rangle\!\rangle-\langle\!\langle C_{\bm k\uparrow}C^\dagger_{\bm p\downarrow}C_{\bm q\downarrow}S^-|C^\dagger_{\bm k'\uparrow}\rangle\!\rangle-2\langle\!\langle C_{\bm k\uparrow}C_{\bm p\downarrow}^\dagger C_{\bm q\uparrow}S^z|C^\dagger_{\bm k'\uparrow}\rangle\!\rangle\Big]$$
\begin{equation}
-\frac{J_{sd}}{2 N_s}\sum_{\bm q}\Big[-\frac{1}{2}\langle\!\langle C_{\bm q\uparrow}S^-|C^\dagger_{\bm k'\uparrow}\rangle\!\rangle\Big],
\label{eq:A7}
\end{equation}

$$(i\omega_n-\xi_{\bm k})\langle\!\langle C_{\bm k\downarrow}S^z|C^\dagger_{\bm k'\uparrow}\rangle\!\rangle=\alpha k_-^2\langle\!\langle C_{\bm k\uparrow}S^z|C^\dagger_{\bm k'\uparrow}\rangle\!\rangle-\frac{J_{sd}}{2 N_s}\sum_{\bm q\bm p}\Big[\langle\!\langle C_{\bm k \downarrow}C^\dagger_{\bm p\downarrow}C_{\bm q\uparrow}S^+|C^\dagger_{\bm k'\uparrow}\rangle\!\rangle-\langle\!\langle C_{\bm k\downarrow}C^\dagger_{\bm p\uparrow}C_{\bm q\downarrow}S^-|C^\dagger_{\bm k'\uparrow}\rangle\!\rangle\Big]$$
\begin{equation}
+\frac{J_{sd}}{2 N_s}\sum_{\bm q}\Big[\langle\!\langle C_{\bm q\downarrow}(S^z)^2|C^\dagger_{\bm k'\uparrow}\rangle\!\rangle+\frac{1}{2}\langle\!\langle C_{\bm q\uparrow}S^+|C^\dagger_{\bm k'\uparrow}\rangle\!\rangle\Big],
\label{eq:A8}
\end{equation}

$$(i\omega_n-\xi_{\bm k})\langle\!\langle C_{\bm k\uparrow}S^+|C^\dagger_{\bm k'\uparrow}\rangle\!\rangle=\alpha k_+^2\langle\!\langle C_{\bm k\downarrow}S^+|C^\dagger_{\bm k'\uparrow}\rangle\!\rangle-\frac{J_{sd}}{2 N_s}\sum_{\bm q\bm p}\Big[-\langle\!\langle C_{\bm k\uparrow}C^\dagger_{\bm p\uparrow}C_{\bm q\uparrow}S^+|C^\dagger_{\bm k'\uparrow}\rangle\!\rangle+\langle\!\langle C_{\bm k\uparrow}C^\dagger_{\bm p\downarrow}C_{\bm q\downarrow}S^+|C^\dagger_{\bm k'\uparrow} \rangle\!\rangle+2\langle\!\langle C_{\bm k\uparrow}C^\dagger_{\bm p\uparrow}C_{\bm q\downarrow}S^z|C^\dagger_{\bm k'\uparrow}\rangle\!\rangle\Big]$$
\begin{equation}
-\frac{J_{sd}}{2 N_s}\sum_{\bm q}\Big[\frac{1}{2}\langle\!\langle C_{\bm q\uparrow}S^+|C^\dagger_{\bm k'\uparrow}\rangle\!\rangle+\frac{3}{4}\langle\!\langle C_{\bm q\downarrow}|C^\dagger_{\bm k'\uparrow}\rangle\!\rangle-\langle\!\langle C_{\bm q\downarrow}S^z|C^\dagger_{\bm k'\uparrow} \rangle\!\rangle-\langle\!\langle C_{\bm q\downarrow}(S^z)^2|C^\dagger_{\bm k'\uparrow} \rangle\!\rangle\Big],
\label{eq:A9}
\end{equation}
and
$$(i\omega_n-\xi_{\bm k})\langle\!\langle C_{\bm k\downarrow}S^+|C^\dagger_{\bm k'\uparrow}\rangle\!\rangle=\alpha k_-^2\langle\!\langle C_{\bm k\uparrow}S^+|C^\dagger_{\bm k'\uparrow}\rangle\!\rangle-\frac{J_{sd}}{2 N_s}\sum_{\bm q\bm p}\Big[-\langle\!\langle C_{\bm k\downarrow}C^\dagger_{\bm p\uparrow}C_{\bm q\uparrow}S^+|C^\dagger_{\bm k'\uparrow}\rangle\!\rangle+\langle\!\langle C_{\bm k\downarrow}C^\dagger_{\bm p\downarrow}C_{\bm q\downarrow}S^+|C^\dagger_{\bm k'\uparrow} \rangle\!\rangle+ 2\langle\!\langle C_{\bm k\downarrow}C^\dagger_{\bm p\uparrow}C_{\bm q\downarrow}S^z|C^\dagger_{\bm k'\uparrow}\rangle\!\rangle\Big]$$
\begin{equation}
-\frac{J_{sd}}{2 N_s}\sum_{\bm q}\Big[-\frac{1}{2}\langle\!\langle C_{\bm q\downarrow}S^+|C^\dagger_{\bm k'\uparrow}\rangle\!\rangle\Big].
\label{eq:A10}
\end{equation}	
}

The Green's functions above have been derived under the assumption $\langle S^z\rangle=0$, which is justified in the absence of a magnetic field and single-ion anisotropy.

\onecolumngrid
\section{Auxiliary Green's functions}
\label{appendix:B}
We first rewrite Eqs.~(\ref{eq:A5})--(\ref{eq:A10}) as follows
\begin{equation}
\langle\!\langle C_{\bm k\downarrow}S^z|C^\dagger_{\bm k'\uparrow}\rangle\!\rangle=-h_{\bm k-}(i\omega_n)X^{(2)}_{\bm k\bm k'}-g_{\bm k}(i\omega_n)X^{(1)}_{\bm k\bm k'},
\label{eq:B1}	
\end{equation}
\begin{equation}
\langle\!\langle C_{\bm k\uparrow}S^z|C^\dagger_{\bm k'\uparrow}\rangle\!\rangle=-h_{\bm k+}(i\omega_n)X^{(1)}_{\bm k\bm k'}-g_{\bm k}(i\omega_n)X^{(2)}_{\bm k\bm k'},
\label{eq:B2}
\end{equation}
\begin{equation}
\langle\!\langle C_{\bm k\downarrow}S^-|C^\dagger_{\bm k'\uparrow}\rangle\!\rangle=-h_{\bm k -}(i\omega_n)Y^{(2)}_{\bm k\bm k'}-g_{\bm k}(i\omega_n)Y^{(1)}_{\bm k\bm k'},
\label{eq:B3}
\end{equation}
\begin{equation}
\langle\!\langle C_{\bm k\uparrow}S^-|C^\dagger_{\bm k'\uparrow}\rangle\!\rangle=-h_{\bm k+}(i\omega_n)Y^{(1)}_{\bm k\bm k'} -g_{\bm k}(i\omega_n)Y^{(2)}_{\bm k\bm k'},
\label{eq:B4}
\end{equation}
\begin{equation}
\langle\!\langle C_{\bm k\downarrow}S^+|C^\dagger_{\bm k'\uparrow}\rangle\!\rangle=-h_{\bm k-}(i\omega_n)Z^{(2)}_{\bm k\bm k'}-g_{\bm k}(i\omega_n)Z^{(1)}_{\bm k\bm k'},
\label{eq:B5}
\end{equation}
\begin{equation}
\langle\!\langle C_{\bm k\uparrow}S^+|C^\dagger_{\bm k'\uparrow}\rangle\!\rangle=-h_{\bm k+}(i\omega_n)Z^{(1)}_{\bm k\bm k'} -g_{\bm k}(i\omega_n)Z^{(2)}_{\bm k\bm k'},
\label{eq:B6}
\end{equation}
where
\begin{equation}
X^{(1)}_{\bm k\bm k'}=\frac{J_{sd}}{2 N_s}\sum_{\bm q\bm p}\Big[\langle\!\langle C_{\bm k \downarrow}C^\dagger_{\bm p\downarrow}C_{\bm q\uparrow}S^+|C^\dagger_{\bm k'\uparrow}\rangle\!\rangle-\langle\!\langle C_{\bm k\downarrow}C^\dagger_{\bm p\uparrow}C_{\bm q\downarrow}S^-|C^\dagger_{\bm k'\uparrow}\rangle\!\rangle\Big]-\frac{J_{sd}}{2 N_s}\sum_{\bm q}\Big[\langle\!\langle C_{\bm q\downarrow}(S^z)^2|C^\dagger_{\bm k'\uparrow}\rangle\!\rangle+\frac{1}{2}\langle\!\langle C_{\bm q\uparrow}S^+|C^\dagger_{\bm k'\uparrow}\rangle\!\rangle\Big],
\label{eq:B7}
\end{equation}
\begin{equation}
X^{(2)}_{\bm k\bm k'}=\frac{J_{sd}}{2 N_s}\sum_{\bm q\bm p}\Big[\langle\!\langle C_{\bm k\uparrow}C^\dagger_{\bm p\downarrow}C_{\bm q\uparrow}S^+|C^\dagger_{\bm k'\uparrow}\rangle\!\rangle-\langle\!\langle C_{\bm k\uparrow}C^\dagger_{\bm p\uparrow}C_{\bm q\downarrow}S^-|C^\dagger_{\bm k'\uparrow}\rangle\!\rangle\Big]+\frac{J_{sd}}{2 N_s}\sum_{\bm q}\Big[\langle\!\langle C_{\bm q\uparrow}(S^z)^2|C^\dagger_{\bm k'\uparrow}\rangle\!\rangle +\frac{1}{2}\langle\!\langle C_{\bm q\downarrow}S^-|C^\dagger_{\bm k'\uparrow}\rangle\!\rangle\Big],
\label{eq:B8}
\end{equation}
$$
Y^{(1)}_{\bm k\bm k'}=\frac{J_{sd}}{2 N_s}\sum_{\bm q\bm p}\Big[\langle\!\langle C_{\bm k\downarrow}C^\dagger_{\bm p\uparrow}C_{\bm q\uparrow}S^-|C^\dagger_{\bm k'\uparrow}\rangle\!\rangle-\langle\!\langle C_{\bm k\downarrow}C^\dagger_{\bm p\downarrow}C_{\bm q\downarrow}S^-|C^\dagger_{\bm k'\uparrow}\rangle\!\rangle-2\langle\!\langle C_{\bm k\downarrow}C^\dagger_{\bm p\downarrow}C_{\bm q\uparrow}S^z|C^\dagger_{\bm k'\uparrow}\rangle\!\rangle\Big]$$
\begin{equation}
+\frac{J_{sd}}{2 N_s} \sum_{\bm q}\Big[\frac{3}{4}\langle\!\langle C_{\bm q\uparrow}|C^\dagger_{\bm k'\uparrow}\rangle\!\rangle-\langle\!\langle C_{\bm q\uparrow}(S^z)^2|C^\dagger_{\bm k'\uparrow}\rangle\!\rangle+\frac{1}{2}\langle\!\langle C_{\bm q\downarrow}S^-|C^\dagger_{\bm k'\uparrow}\rangle\!\rangle+\langle\!\langle C_{\bm q\uparrow}S^z|C^\dagger_{\bm k'\uparrow}\rangle\!\rangle\Big],
\label{eq:B9}
\end{equation}

\begin{equation}
Y^{(2)}_{\bm k\bm k'}=\frac{J_{sd}}{2 N_s}\sum_{\bm q\bm p}\Big[\langle\!\langle C_{\bm k\uparrow}C^\dagger_{\bm p\uparrow}C_{\bm q\uparrow}S^-|C^\dagger_{\bm k'\uparrow}\rangle\!\rangle-\langle\!\langle C_{\bm k\uparrow}C^\dagger_{\bm p\downarrow}C_{\bm q\downarrow}S^-|C^\dagger_{\bm k'\uparrow}\rangle\!\rangle-2\langle\!\langle C_{\bm k\uparrow}C_{\bm p\downarrow}^\dagger C_{\bm q\uparrow}S^z|C^\dagger_{\bm k'\uparrow}\rangle\!\rangle\Big]+\frac{J_{sd}}{2 N_s}\sum_{\bm q}\Big[-\frac{1}{2}\langle\!\langle C_{\bm q\uparrow}S^-|C^\dagger_{\bm k'\uparrow}\rangle\!\rangle\Big],
\label{eq:B10}
\end{equation}

\begin{equation}
Z^{(1)}_{\bm k\bm k'}=\frac{J_{sd}}{2 N_s}\sum_{\bm q\bm p}\Big[-\langle\!\langle C_{\bm k\downarrow}C^\dagger_{\bm p\uparrow}C_{\bm q\uparrow}S^+|C^\dagger_{\bm k'\uparrow}\rangle\!\rangle+\langle\!\langle C_{\bm k\downarrow}C^\dagger_{\bm p\downarrow}C_{\bm q\downarrow}S^+|C^\dagger_{\bm k'\uparrow} \rangle\!\rangle+ 2\langle\!\langle C_{\bm k\downarrow}C^\dagger_{\bm p\uparrow}C_{\bm q\downarrow}S^z|C^\dagger_{\bm k'\uparrow}\rangle\!\rangle\Big]+\frac{J_{sd}}{2 N_s}\sum_{\bm q}\Big[-\frac{1}{2}\langle\!\langle C_{\bm q\downarrow}S^+|C^\dagger_{\bm k'\uparrow}\rangle\!\rangle\Big],
\label{eq:B11}
\end{equation}
$$Z^{(2)}_{\bm k\bm k'}=\frac{J_{sd}}{2N_s}\sum_{\bm q\bm p}\Big[-\langle\!\langle C_{\bm k\uparrow}C^\dagger_{\bm p\uparrow}C_{\bm q\uparrow}S^+|C^\dagger_{\bm k'\uparrow}\rangle\!\rangle+\langle\!\langle C_{\bm k\uparrow}C^\dagger_{\bm p\downarrow}C_{\bm q\downarrow}S^+|C^\dagger_{\bm k'\uparrow} \rangle\!\rangle+2\langle\!\langle C_{\bm k\uparrow}C^\dagger_{\bm p\uparrow}C_{\bm q\downarrow}S^z|C^\dagger_{\bm k'\uparrow}\rangle\!\rangle\Big]$$
\begin{equation}
+\frac{J_{sd}}{2 N_s}\sum_{\bm q}\Big[\frac{1}{2}\langle\!\langle C_{\bm q\uparrow}S^+|C^\dagger_{\bm k'\uparrow}\rangle\!\rangle+\frac{3}{4}\langle\!\langle C_{\bm q\downarrow}|C^\dagger_{\bm k'\uparrow}\rangle\!\rangle-\langle\!\langle C_{\bm q\downarrow}S^z|C^\dagger_{\bm k'\uparrow} \rangle\!\rangle-\langle\!\langle C_{\bm q\downarrow}(S^z)^2|C^\dagger_{\bm k'\uparrow} \rangle\!\rangle\Big].
\label{eq:B12}
\end{equation}

To proceed, we first need to decouple the higher-order Green's functions appearing in Eqs.~(\ref{eq:B7})--(\ref{eq:B12}). To this end, we follow the procedure of Nagaoka \cite{nagaoka1965,yanagisawa2012,yanagisawa2015}, truncating the hierarchy of EOM by expressing higher-order Green's functions in terms of lower-order ones, while retaining only averages of the form $\langle C^\dagger_{\bm k'\uparrow}C_{\bm k\downarrow}S^-\rangle$, $\langle C^\dagger_{\bm k'\downarrow}C_{\bm k\uparrow}S^+\rangle$, and $\langle C^\dagger_{\bm k'\sigma}C_{\bm k\sigma}S^z\rangle$, which conserve the total spin projection along the $z$-axis. For instance, in the correlator $\langle C^\dagger_{\bm k'\uparrow}C_{\bm k\downarrow}S^-\rangle$, the impurity operator $S^-$ lowers the impurity spin by $\Delta S^z=-1$, while the electronic part $C^\dagger_{\bm k'\uparrow}C_{\bm k\downarrow}$ raises the conduction-electron spin by $\Delta s^z=+1$. Therefore, the total $z$-component of spin is conserved.  

Using the above decoupling procedure, we obtain

\begin{equation}
\langle\!\langle C_{\bm k\uparrow}C_{\bm p\uparrow}^\dagger C_{\bm q\downarrow}S^-|C_{\bm k'\uparrow}^\dagger \rangle\!\rangle
\simeq\langle C_{\bm k\uparrow}C^\dagger_{\bm p\uparrow}\rangle\langle\!\langle C_{\bm q\downarrow}S^-|C_{\bm k'\uparrow}^\dagger\rangle\!\rangle
+\langle  C^\dagger_{\bm p\uparrow}C_{\bm q\downarrow}S^-\rangle\langle\!\langle C_{\bm k\uparrow}|C_{\bm k'\uparrow}^\dagger\rangle\!\rangle,
\label{eq:B13}
\end{equation}
\begin{equation}
\langle\!\langle C_{\bm k\uparrow}C_{\bm p\downarrow}^\dagger C_{\bm q\uparrow}S^+|C_{\bm k'\uparrow}^\dagger\rangle\!\rangle\simeq
\langle C^\dagger_{\bm p\downarrow}C_{\bm q\uparrow}S^+\rangle\langle\!\langle C_{\bm k\uparrow}|C_{\bm k'\uparrow}^\dagger\rangle\!\rangle-\langle C^\dagger_{\bm p\downarrow}C_{\bm k\uparrow}S^+\rangle\langle\!\langle C_{\bm q\uparrow}|C_{\bm k'\uparrow}^\dagger\rangle\!\rangle,
\label{eq:B14}
\end{equation}
\begin{equation}
\langle\!\langle C_{\bm k\downarrow}C^\dagger_{\bm p\downarrow}C_{\bm q\uparrow}S^z|C_{\bm k'\uparrow}^\dagger\rangle\!\rangle
\simeq \langle C_{\bm k\downarrow}C^\dagger_{\bm p\downarrow}\rangle\langle\!\langle C_{\bm q\uparrow}S^z|C_{\bm k'\uparrow}^\dagger\rangle\!\rangle-\langle C^\dagger_{\bm p\downarrow}C_{\bm k\downarrow}S^z\rangle\langle\!\langle C_{\bm q\uparrow}|C_{\bm k'\uparrow}^\dagger\rangle\!\rangle,
\label{eq:B15}
\end{equation}
\begin{equation}
\langle\!\langle C_{\bm k\downarrow}C_{\bm p\downarrow}^\dagger C_{\bm q\downarrow} S^-|C_{\bm k'\uparrow}^\dagger\rangle\!\rangle
\simeq\langle C_{\bm p\downarrow}^\dagger C_{\bm q \downarrow}\rangle\langle\!\langle C_{\bm k\downarrow} S^-|C_{\bm k'\uparrow}^\dagger\rangle\!\rangle+\langle C_{\bm k\downarrow}C^\dagger_{\bm p\downarrow}\rangle\langle\!\langle C_{\bm q\downarrow}S^-|C_{\bm k'\uparrow}^\dagger\rangle\!\rangle,
\label{eq:B16}
\end{equation}
\begin{equation}
\langle\!\langle C_{\bm k\downarrow}C^\dagger_{\bm p\uparrow}C_{\bm q\uparrow}S^-|C_{\bm k'\uparrow}^\dagger\rangle\!\rangle
\simeq \langle C_{\bm p\uparrow}^\dagger C_{\bm q\uparrow}\rangle\langle\!\langle C_{\bm k\downarrow} S^-|C_{\bm k'\uparrow}^\dagger\rangle\!\rangle-\langle C_{\bm p\uparrow}^\dagger C_{\bm k\downarrow}S^-\rangle\langle\!\langle C_{\bm q\uparrow}|C_{\bm k'\uparrow}^\dagger\rangle\!\rangle,
\label{eq:B17}
\end{equation}
\begin{equation}
\langle\!\langle C_{\bm k\downarrow}C_{\bm p\downarrow}^\dagger C_{\bm q\uparrow}S^+|C_{\bm k'\uparrow}^\dagger \rangle\!\rangle
\simeq \langle C_{\bm k\downarrow}C^\dagger_{\bm p\downarrow}\rangle\langle\!\langle C_{\bm q\uparrow}S^+|C^\dagger_{\bm k'\uparrow}\rangle\!\rangle+\langle C^\dagger_{\bm p\downarrow}C_{\bm q\uparrow}S^+\rangle \langle\!\langle C_{\bm k\downarrow}|C^\dagger_{\bm k'\uparrow}\rangle\!\rangle,
\label{eq:B18}
\end{equation}
\begin{equation}
\langle\!\langle C_{\bm k\uparrow}C_{\bm p\uparrow}^\dagger C_{\bm q\uparrow}S^+|C_{\bm k'\uparrow}^\dagger\rangle\!\rangle
\simeq\langle C_{\bm k\uparrow}C^\dagger_{\bm p\uparrow}\rangle\langle\!\langle C_{\bm q\uparrow}S^+|C^\dagger_{\bm k'\uparrow}\rangle\!\rangle+\langle C^\dagger_{\bm p\uparrow}C_{\bm q\uparrow}\rangle \langle\!\langle C_{\bm k\uparrow}S^+|C^\dagger_{\bm k'\uparrow}\rangle\!\rangle,
\label{eq:B19}
\end{equation}
\begin{equation}
\langle\!\langle C_{\bm k\uparrow}C^\dagger_{\bm p\downarrow}C_{\bm q\downarrow}S^+|C_{\bm k'\uparrow}^\dagger\rangle\!\rangle
\simeq \langle C^\dagger_{\bm p\downarrow}C_{\bm q\downarrow}\rangle \langle\!\langle C_{\bm k\uparrow}S^+|C^\dagger_{\bm k'\uparrow}\rangle\!\rangle-\langle C^\dagger_{\bm p\downarrow}C_{\bm k\uparrow}S^+\rangle \langle\!\langle C_{\bm q\downarrow}|C^\dagger_{\bm k'\uparrow}\rangle\!\rangle,
\label{eq:B20}
\end{equation}
\begin{equation}
\langle\!\langle C_{\bm k\downarrow}C_{\bm p\uparrow}^\dagger C_{\bm q\downarrow} S^-|C_{\bm k'\uparrow}^\dagger\rangle\!\rangle\simeq 
\langle C^\dagger_{\bm p\uparrow}C_{\bm q\downarrow}S^-\rangle\langle\!\langle C_{\bm k\downarrow}|C^\dagger_{\bm k'\uparrow}\rangle\!\rangle -\langle C^\dagger_{\bm p\uparrow}C_{\bm k\downarrow}S^-\rangle \langle\!\langle C_{\bm q\downarrow}|C^\dagger_{\bm k'\uparrow}\rangle\!\rangle,
\label{eq:B21}
\end{equation}
\begin{equation}
\langle\!\langle C_{\bm k\uparrow}C^\dagger_{\bm p\uparrow}C_{\bm q\downarrow}S^z|C_{\bm k'\uparrow}^\dagger\rangle\!\rangle
\simeq \langle C_{\bm k\uparrow}C^\dagger_{\bm p\uparrow}\rangle\langle\!\langle C_{\bm q\downarrow}S^z|C^\dagger_{\bm k'\uparrow}\rangle\!\rangle-\langle C^\dagger_{\bm p\uparrow}C_{\bm k\uparrow}S^z \rangle \langle\!\langle C_{\bm q\downarrow}|C^\dagger_{\bm k'\uparrow}\rangle\!\rangle.
\label{eq:B22}
\end{equation}
In Eqs.~(\ref{eq:B13})--(\ref{eq:B22}), we assume that $\langle C^\dagger_{\bm k\uparrow}C_{\bm p\downarrow}\rangle\simeq 0$, because this off-diagonal spin correlator is linear in $\alpha$ in the weak spin-splitting regime and can therefore be neglected, see Sec.~\ref{sec:4}.

In addition to the above decouplings, we have also used $\langle\!\langle C_{\bm k\sigma}(S^z)^2|C^\dagger_{\bm k'\sigma'}\rangle\!\rangle\simeq\langle (S^z)^2\rangle \langle\!\langle C_{\bm k\sigma}|C^\dagger_{\bm k'\sigma'}\rangle\!\rangle$. We note that $\langle (S^z)^2\rangle$ remains finite even in the absence of magnetic field and single-ion anisotropy, taking the value $\langle (S^z)^2\rangle=1/4$. The statistical average is defined as $\langle \hdots\rangle={\rm Tr}({\rm exp}(-\beta \mathcal H)\hdots)/{\rm Tr}({\rm exp}(-\beta\mathcal H))$.

Further, we introduce the quantities $n_{\bm k}$ and $m_{\bm k}$ as follows
\vspace{2mm}
\begin{equation}
n_{\bm k}=\sum_{\bm p}\langle C^\dagger_{\bm p\uparrow}C_{\bm k\uparrow}
\rangle=\sum_{\bm p}\langle C^\dagger_{\bm p\downarrow}C_{\bm k\downarrow}\rangle,
\label{eq:B23}
\end{equation}
and
\vspace{-3mm}
\begin{equation}
m_{\bm k}/3=\sum_{\bm p}\langle C^\dagger_{\bm p\uparrow}C_{\bm k\downarrow}S^-\rangle=\sum_{\bm p}\langle C^\dagger_{\bm p\downarrow}C_{\bm k\uparrow}S^+\rangle
=2\sum_{\bm p}\langle C^\dagger_{\bm p\uparrow}C_{\bm k\uparrow}S^z\rangle=-2\sum_{\bm p}\langle C^\dagger_{\bm p\downarrow}C_{\bm k\downarrow}S^z\rangle.
\label{eq:B24}
\end{equation}

Using the decouplings in Eqs.~(\ref{eq:B13})--(\ref{eq:B22}), we obtain the explicit expressions for Eqs.~(\ref{eq:B7})--(\ref{eq:B12}) as follows
\begin{equation}
X^{(1)}_{\bm k\bm k'}=-\frac{J_{sd}}{2N_s}\left(n_{\bm k}-\frac{1}{2}\right)\sum_{\bm q}\langle\!\langle C_{\bm q\uparrow}S^+|C^\dagger_{\bm k'\uparrow}\rangle\!\rangle+\frac{J_{sd}}{2N_s}\left(\frac{m_{\bm k}}{3}-\langle \left(S^z\right)^2\rangle\right) \sum_{\bm q}\mathcal F_{\bm q\bm k'},
\label{eq:B25}
\end{equation}
\begin{equation}
X^{(2)}_{\bm k\bm k'}=\frac{J_{sd}}{2N_s}\left(\langle \left(S^z\right)^2\rangle-\frac{m_{\bm k}}{3}\right) \sum_{\bm q}\mathcal G_{\bm q\bm k'}+\frac{J_{sd}}{2N_s}\left(n_{\bm k}-\frac{1}{2}\right)\sum_{\bm q}\langle\!\langle C_{\bm q\downarrow}S^-|C^\dagger_{\bm k'\uparrow}\rangle\!\rangle,
\label{eq:B26}
\end{equation}
\begin{equation}
Y^{(1)}_{\bm k\bm k'}=\frac{J_{sd}}{2N_s}\left(\frac{3}{4}-\langle (S^z)^2\rangle-\frac{2}{3}m_{\bm  k}\right)\sum_{\bm q}\mathcal G_{\bm q\bm k'}+\frac{J_{sd}}{2N_s}\left(n_{\bm k}-\frac{1}{2}\right)\sum_{\bm q}\langle\!\langle C_{\bm q\downarrow}S^-|C^\dagger_{\bm k'\uparrow}\rangle\!\rangle+\frac{J_{sd}}{N_s}\left(n_{\bm k}-\frac{1}{2}\right)\sum_{\bm q}\langle\!\langle C_{\bm q\uparrow}S^z|C^\dagger_{\bm k'\uparrow}\rangle\!\rangle,
\label{eq:B27}
\end{equation}
\begin{equation}
Y^{(2)}_{\bm k\bm k'}=-\frac{J_{sd}}{2N_s}\left(n_{\bm k}-\frac{1}{2}\right)\sum_{\bm q}\langle\!\langle C_{\bm q\uparrow}S^-|C^\dagger_{\bm k'\uparrow}\rangle\!\rangle,
\label{eq:B28}
\end{equation}
\begin{equation}
Z^{(1)}_{\bm k\bm k'}=-\frac{J_{sd}}{2N_s}\left(n_{\bm k}-\frac{1}{2}\right)\sum_{\bm q}\langle\!\langle C_{\bm q\downarrow}S^+|C^\dagger_{\bm k'\uparrow}\rangle\!\rangle,
\label{eq:B29}
\end{equation}
\begin{equation}
Z^{(2)}_{\bm k\bm k'}=\frac{J_{sd}}{2N_s}\left(\frac{3}{4}-\langle (S^z)^2 \rangle-\frac{2}{3}m_{\bm k}\right)\sum_{\bm q}\mathcal F_{\bm q\bm k'}+\frac{J_{sd}}{2N_s}\left(n_{\bm k}-\frac{1}{2}\right)\sum_{\bm q}\langle\!\langle C_{\bm q\uparrow}S^+|C^\dagger_{\bm k'\uparrow}\rangle\!\rangle-\frac{J_{sd}}{N_s}\left(n_{\bm k}-\frac{1}{2}\right)\sum_{\bm q}\langle\!\langle C_{\bm q\downarrow}S^z|C^\dagger_{\bm k'\uparrow}\rangle\!\rangle.
\label{eq:B30}
\end{equation}
\section{Derivation of Eqs.~(\ref{eq:14a}) and (\ref{eq:14b})}
\label{appendix:C}
In Eqs.~(\ref{eq:10}) and (\ref{eq:11}), we introduced the following functions

$$\Gamma_{\bm k\bm k'}=\langle\!\langle C_{\bm k\uparrow}S^z|C^\dagger_{\bm k'\uparrow}\rangle\!\rangle+\langle\!\langle C_{\bm k\downarrow}S^-|C^\dagger_{\bm k'\uparrow}\rangle\!\rangle$$
\begin{equation}=-\frac{i\omega_n-\xi_{\bm k}}{(i\omega_n-\xi_{\bm k})^2-\alpha^2(k_x^2+k_y^2)^2}\left(X_{\bm k\bm k'}^{(2)}+Y_{\bm k\bm k'}^{(1)}\right)+\frac{1}{(i\omega_n-\xi_{\bm k})^2-\alpha^2(k_x^2+k_y^2)^2}\left(-\alpha k_-^2 Y_{\bm k\bm k'}^{(2)}- \alpha k_+^2 X_{\bm k\bm k'}^{(1)}\right),
\label{eq:C1}	
\end{equation}
and
\begin{equation*}
\Delta_{\bm k\bm k'}=\langle\!\langle C_{\bm k\downarrow}S^z|C^\dagger_{\bm k'\uparrow}\rangle\!\rangle-\langle\!\langle C_{\bm k\uparrow}S^+|C^\dagger_{\bm k'\uparrow}\rangle\!\rangle
\end{equation*}
\begin{equation}=\frac{i\omega_n-\xi_{\bm k}}{(i\omega_n-\xi_{\bm k})^2-\alpha^2(k_x^2+k_y^2)^2}\Big(Z_{\bm k\bm k'}^{(2)}-X_{\bm k\bm k'}^{(1)}\Big)
+\frac{1}{(i\omega_n-\xi_{\bm k})^2-\alpha^2(k_x^2+k_y^2)^2}\left(-\alpha k_-^2 X_{\bm k\bm k'}^{(2)}+\alpha k_+^2 Z_{\bm k\bm k'}^{(1)}\right).
\label{eq:C2}
\end{equation}

Substituting the decoupled expressions for $X^{(1,2)}_{\bm k\bm k'},\, Y^{(1,2)}_{\bm k\bm k'},\, Z^{(1,2)}_{\bm k\bm k'}$, derived in Appendix~\ref{appendix:B}, into Eqs.~(\ref{eq:C1}) and (\ref{eq:C2}) and summing over $\bm k$, we obtain 
\begin{equation}\mathcal E(i\omega_n)\sum_{\bm q}\Gamma_{\bm q\bm k'}=-\mathcal O(i\omega_n)\sum_{\bm q}\mathcal G_{\bm q\bm k'}-\mathcal Q_+(i\omega_n)\sum_{\bm q}\mathcal F_{\bm q\bm k'}
-\mathcal L_+(i\omega_n)\sum_{\bm q}\langle\!\langle C_{\bm q\uparrow}S^+|C^\dagger_{\bm k'\uparrow}\rangle\!\rangle-\mathcal L_-(i\omega_n)\sum_{\bm q}\langle\!\langle C_{\bm q\uparrow}S^-|C^\dagger_{\bm k'\uparrow}\rangle\!\rangle,
\label{eq:C3}
\end{equation}
\begin{equation}\mathcal E(i\omega_n)\sum_{\bm q}\Delta_{\bm q\bm k'}=\mathcal Q_-(i\omega_n)\sum_{\bm q}\mathcal G_{\bm q\bm k'}+\mathcal O(i\omega_n)\sum_{\bm q}\mathcal F_{\bm q\bm k'}
+\mathcal L_-(i\omega_n)\sum_{\bm q}\langle\!\langle C_{\bm q\downarrow}S^-|C^\dagger_{\bm k'\uparrow}\rangle\!\rangle+\mathcal L_+(i\omega_n)\sum_{\bm q}\langle\!\langle C_{\bm q\downarrow}S^+|C^\dagger_{\bm k'\uparrow}\rangle\!\rangle,
\label{eq:C4}
\end{equation}
where

\begin{subequations}
\begin{equation} 
\mathcal E(i\omega_n)=1+\frac{J_{sd}}{N_s}\sum_{\bm k}\frac{i\omega_n-\xi_{\bm k}}{(i\omega_n-\xi_{\bm k})^2-\alpha^2k^4}\left(n_{\bm k}-\frac{1}{2}\right),	\quad \mathcal O(i\omega_n)=\frac{J_{sd}}{2N_s}\sum_{\bm k}\frac{i\omega_n-\xi_{\bm k}}{(i\omega_n-\xi_{\bm k})^2-\alpha^2k^4}\left(\frac{3}{4}-m_{\bm k}\right),	
\label{eq:C5a}
\end{equation}

\begin{equation}
\mathcal L_\pm(i\omega_n)=-\frac{J_{sd}}{2N_s}\sum_{\bm k}\frac{\alpha k_\pm^2 }{(i\omega_n-\xi_{\bm k})^2-\alpha^2k^4}\left(n_{\bm k}-\frac{1}{2}\right),	\quad  \mathcal Q_\pm(i\omega_n)=-\frac{J_{sd}}{2N_s}\sum_{\bm k}\frac{\alpha k_\pm^2}{(i\omega_n-\xi_{\bm k})^2-\alpha^2k^4}\left(\frac{1}{4}-\frac{m_{\bm k}}{3}\right).
\label{eq:C5b}
\end{equation}
\end{subequations}

In Eqs.~(\ref{eq:C5a}) and (\ref{eq:C5b}), $\mathcal E(i\omega_n)$ and $\mathcal O(i\omega_n)$ arise from $s$-$d$ exchange scattering and describe the influence of the localized moments on the itinerant electrons. By contrast, $\mathcal L(i\omega_n)$ and $\mathcal Q(i\omega_n)$ originate from the combined action of quadratic spin splitting and the $s$-$d$ exchange interaction, thereby capturing the interplay between the spin-split conduction band and impurity scattering.

Substituting $\sum_{\bm q}\Gamma_{\bm q\bm k'}$ and $\sum_{\bm q}\Delta_{\bm q\bm k'}$ into Eqs.~(\ref{eq:10}) and (\ref{eq:11}) and solving the resulting system, we obtain
\begin{equation}\mathcal G_{\bm k\bm k'}(i\omega_n)=\frac{i\omega_n-\xi_{\bm k}}{(i\omega_n-\xi_{\bm k})^2-\alpha^2(k_x^2+k_y^2)^2}\delta_{\bm k\bm k'}
+\mathcal U_{\bm k}(i\omega_n)\sum_{\bm q}\mathcal G_{\bm q\bm k'}+\mathcal R_{\bm k}(i\omega_n)\sum_{\bm q}\mathcal F_{\bm q\bm k'},
\label{eq:C6}
\end{equation}
and
\begin{equation}
\mathcal F_{\bm k\bm k'}(i\omega_n)=\frac{\alpha k_-^2}{(i\omega_n-\xi_{\bm k})^2-\alpha^2(k_x^2+k_y^2)^2}\delta_{\bm k\bm k'}
+\mathcal I_{\bm k}(i\omega_n)\sum_{\bm q}\mathcal G_{\bm q\bm k'}+\mathcal B_{\bm k}(i\omega_n)\sum_{\bm q}\mathcal F_{\bm q\bm k'}.
\label{eq:C7}
\end{equation}

For convenience, in Eqs.~(\ref{eq:C6}) and (\ref{eq:C7}) the combinations of the functions introduced in Eqs.~(\ref{eq:C5a})--(\ref{eq:C5b}) have been grouped into four auxiliary functions $\mathcal U_{\bm k},\mathcal R_{\bm k},\mathcal I_{\bm k}$, and $\mathcal B_{\bm k}$, defined as follows 

\begin{subequations}
\begin{equation}
\mathcal U_{\bm k}(i\omega_n)=\frac{J_{sd}}{2N_s}\frac{1}{\mathcal E(i\omega_n)}\Big[ \Xi(i\omega_n)g_{\bm k}(i\omega_n)+ \Pi(i\omega_n)h_{\bm k+}(i\omega_n)\Big],
\label{eq:C8a}
\end{equation}	

\begin{equation}
\mathcal R_{\bm k}(i\omega_n)=\frac{J_{sd}}{2N_s}\frac{1}{\mathcal E(i\omega_n)}\Big[ \Phi(i\omega_n)g_{\bm k}(i\omega_n)+\Omega(i\omega_n)h_{\bm k+}(i\omega_n)\Big],
\label{eq:C8b}
\end{equation}

\begin{equation}
\mathcal I_{\bm k}(i\omega_n)=\frac{J_{sd}}{2N_s}\frac{1}{\mathcal E(i\omega_n)}\Big[\Pi(i\omega_n)g_{\bm k}(i\omega_n)+ \Xi(i\omega_n)h_{\bm k-}(i\omega_n)\Big],
\label{eq:C8c}
\end{equation}
\begin{equation}
\mathcal B_{\bm k}(i\omega_n)=\frac{J_{sd}}{2N_s}\frac{1}{\mathcal E(i\omega_n)}\Big[\Omega(i\omega_n)g_{\bm k}(i\omega_n)+\mathcal \mathcal \Phi(i\omega_n) h_{\bm k-}(i\omega_n)\Big],
\label{eq:C8d}	
\end{equation}
\end{subequations}
where
\begin{subequations}
\begin{equation}\Xi(i\omega_n)=\mathcal L_+(i\omega_n) \frac{\mathcal V_6(i\omega_n)}{\mathcal M_6(i\omega_n)}+\mathcal L_-(i\omega_n) \frac{\mathcal V_4(i\omega_n)}{\mathcal M_4(i\omega_n)}+\mathcal O(i\omega_n),
\label{eq:C9a}
\end{equation}
\begin{equation}\Pi(i\omega_n)=\mathcal L_+(i\omega_n) \frac{\mathcal V_5(i\omega_n)}{\mathcal M_5(i\omega_n)}+\mathcal L_-(i\omega_n) \frac{\mathcal V_3(i\omega_n)}{\mathcal M_3(i\omega_n)}+\mathcal Q_-(i\omega_n),
\label{eq:C9b}
\end{equation}
\begin{equation}\Phi(i\omega_n)=\mathcal L_+(i\omega_n) \frac{\mathcal W_6(i\omega_n)}{\mathcal M_6(i\omega_n)}+\mathcal L_-(i\omega_n) \frac{\mathcal W_4(i\omega_n)}{\mathcal M_4(i\omega_n)}+\mathcal Q_+(i\omega_n),
\label{eq:C9c}
\end{equation}
\begin{equation}\Omega(i\omega_n)=\mathcal L_+(i\omega_n) \frac{\mathcal W_5(i\omega_n)}{\mathcal M_5(i\omega_n)}+\mathcal L_-(i\omega_n) \frac{\mathcal W_3(i\omega_n)}{\mathcal M_3(i\omega_n)}+\mathcal O(i\omega_n).
\label{eq:C9d}
\end{equation}
\end{subequations}
The derivations of the functions $\mathcal W_m(i\omega_n)$ and $\mathcal V_m(i\omega_n)$ are presented in Appendix~\ref{appendix:D}.

By summing Eqs.~(\ref{eq:C6}) and (\ref{eq:C7}) and solving the resulting integral equations, we obtain Eqs.~(\ref{eq:14a}) and (\ref{eq:14b}) in the main text.

\section{Derivation of the functions $\mathcal V_m(i\omega_n)$ and $\mathcal W_m(i\omega_n)$}
\label{appendix:D}

By substituting Eqs.~(\ref{eq:B25})--(\ref{eq:B30}) into Eqs.~(\ref{eq:B1})--(\ref{eq:B6}) and summing both sides of the resulting equations over $\bm k$, we obtain the following system of equations
{
\begin{equation}
\sum_{\bm q}\langle\!\langle C_{\bm q\downarrow}S^z|C^\dagger_{\bm k'\uparrow}\rangle\!\rangle=n_{h-} \sum_{\bm q}\langle\!\langle C_{\bm q\downarrow}S^-|C^\dagger_{\bm k'\uparrow}\rangle\!\rangle+n_g \sum_{\bm q}\langle\!\langle C_{\bm q\uparrow}S^+|C^\dagger_{\bm k'\uparrow}\rangle\!\rangle+S_{h-} \sum_{\bm q}\mathcal G_{\bm q\bm k'}+S_g \sum_{\bm q}\mathcal F_{\bm q\bm k'},
\label{eq:D1}
\end{equation}

\begin{equation}
\sum_{\bm q}\langle\!\langle C_{\bm q\uparrow}S^z|C^\dagger_{\bm k'\uparrow}\rangle\!\rangle=-n_{h+} \sum_{\bm q}\langle\!\langle C_{\bm q\uparrow}S^+|C^\dagger_{\bm k'\uparrow}\rangle\!\rangle-n_g \sum_{\bm q}\langle\!\langle C_{\bm q\downarrow}S^-|C^\dagger_{\bm k'\uparrow}\rangle\!\rangle-S_g\sum_{\bm q}\mathcal G_{\bm q\bm k'}-S_{h+} \sum_{\bm q}\mathcal F_{\bm q\bm k'},
\label{eq:D2}
\end{equation}

\begin{equation}
\Big(1+n_g\Big)\sum_{\bm q}\langle\!\langle C_{\bm q\downarrow}S^-|C^\dagger_{\bm k'\uparrow}\rangle\!\rangle=-n_{h-} \sum_{\bm q}\langle\!\langle C_{\bm q\uparrow}S^-|C^\dagger_{\bm k'\uparrow}\rangle\!\rangle-2n_g \sum_{\bm q}\langle\!\langle C_{\bm q\uparrow}S^z|C^\dagger_{\bm k'\uparrow}\rangle\!\rangle-K_g\sum_{\bm q}\mathcal G_{\bm q\bm k'},
\label{eq:D3}
\end{equation}

\begin{equation}
\Big(1-n_g\Big)\sum_{\bm q}\langle\!\langle C_{\bm q\uparrow}S^-|C^\dagger_{\bm k'\uparrow}\rangle\!\rangle=n_{h+}\sum_{\bm q}\langle\!\langle C_{\bm q\downarrow}S^-|C^\dagger_{\bm k'\uparrow}\rangle\!\rangle+2n_{h+}\sum_{\bm q}\langle\!\langle C_{\bm q\uparrow}S^z|C^\dagger_{\bm k'\uparrow}\rangle\!\rangle+K_{h+}\sum_{\bm q}\mathcal G_{\bm q\bm k'},
\label{eq:D4}
\end{equation}

\begin{equation}
\Big(1-n_g\Big)\sum_{\bm q}\langle\!\langle C_{\bm q\downarrow}S^+|C^\dagger_{\bm k'\uparrow}\rangle\!\rangle=n_{h-} \sum_{\bm q}\langle\!\langle C_{\bm q\uparrow}S^+|C^\dagger_{\bm k'\uparrow}\rangle\!\rangle-2n_{h-} \sum_{\bm q}\langle\!\langle C_{\bm q\downarrow}S^z|C^\dagger_{\bm k'\uparrow}\rangle\!\rangle+K_{h-}\sum_{\bm q}\mathcal F_{\bm q\bm k'},
\label{eq:D5}
\end{equation}

\begin{equation}
\Big(1+n_g\Big)\sum_{\bm q}\langle\!\langle C_{\bm q\uparrow}S^+|C^\dagger_{\bm k'\uparrow}\rangle\!\rangle=-2n_g\sum_{\bm q}\langle\!\langle C_{\bm q\downarrow}S^z|C^\dagger_{\bm k'\uparrow}\rangle\!\rangle-n_{h+}\sum_{\bm q}\langle\!\langle C_{\bm q\downarrow}S^+|C^\dagger_{\bm k'\uparrow}\rangle\!\rangle-K_g\sum_{\bm q}\mathcal F_{\bm q\bm k'},
\label{eq:D6}
\end{equation}
}

where, for brevity, we have introduced

\begin{equation}
n_g=\frac{J_{sd}}{2N_s}\sum_{\bm k}g_{\bm k}\left(n_{\bm k}-\frac{1}{2}\right),\qquad n_{h\pm}=-\frac{J_{sd}}{2N_s}\sum_{\bm k}h_{\bm k\pm}\left(n_{\bm k}-\frac{1}{2}\right), 
\label{eq:D7}
\end{equation}

\begin{equation}
S_g=\frac{J_{sd}}{2N_s}\sum_{\bm k}g_{\bm k}\left(\langle (S^z)^2\rangle-\frac{m_{\bm k}}{3}\right),\qquad S_{h\pm}=-\frac{J_{sd}}{2N_s}\sum_{\bm k}h_{\bm k\pm}\left(\langle (S^z)^2\rangle-\frac{m_{\bm k}}{3}\right),
\label{eq:D8}
\end{equation}

\begin{equation}
K_g=\frac{J_{sd}}{2N_s}\sum_{\bm k}g_{\bm k}\left(\frac{3}{4}-\langle (S^z)^2\rangle-\frac{2}{3}m_{\bm k}\right),\qquad K_{h\pm}=-\frac{J_{sd}}{2N_s}\sum_{\bm k}h_{\bm k\pm}\left(\frac{3}{4}-\langle (S^z)^2\rangle-\frac{2}{3}m_{\bm k}\right).
\label{eq:D9}
\end{equation}

It is seen that solutions of Eqs.~(\ref{eq:D1})--(\ref{eq:D6}) form a closed set of expressions involving the combinations $\sum_{\bm q}\mathcal G_{\bm q\bm k'}$ and $\sum_{\bm q}\mathcal F_{\bm q\bm k'}$, with the corresponding prefactors defining the functions $\mathcal V_m(i\omega_n)$ and $\mathcal W_m(i\omega_n)$. We can therefore rewrite these equations in the form 
{\small
\begin{equation}
\mathcal M_1 \sum_{\bm q}\langle\!\langle C_{\bm q\downarrow}S^z|C^\dagger_{\bm k'\uparrow}\rangle\!\rangle=\mathcal V_1 \sum_{\bm q}\mathcal G_{\bm q\bm k'}+\mathcal W_1\sum_{\bm q}\mathcal F_{\bm q\bm k'},
\label{eq:D11}
\end{equation}
\begin{equation}
\mathcal M_2 \sum_{\bm q}\langle\!\langle C_{\bm q\uparrow}S^z|C^\dagger_{\bm k'\uparrow}\rangle\!\rangle=\mathcal V_2 \sum_{\bm q}\mathcal G_{\bm q\bm k'}+\mathcal W_2\sum_{\bm q}\mathcal F_{\bm q\bm k'},
\label{eq:D12}
\end{equation}
\begin{equation}
\mathcal M_3 \sum_{\bm q}\langle\!\langle C_{\bm q\downarrow}S^-|C^\dagger_{\bm k'\uparrow}\rangle\!\rangle=\mathcal V_3 \sum_{\bm q}\mathcal G_{\bm q\bm k'}+\mathcal W_3\sum_{\bm q}\mathcal F_{\bm q\bm k'},
\label{eq:D13}
\end{equation}
\begin{equation}
\mathcal M_4 \sum_{\bm q}\langle\!\langle C_{\bm q\uparrow}S^-|C^\dagger_{\bm k'\uparrow}\rangle\!\rangle =\mathcal V_4 \sum_{\bm q}\mathcal G_{\bm q\bm k'}+\mathcal W_4\sum_{\bm q}\mathcal F_{\bm q\bm k'},
\label{eq:D14}
\end{equation}
\begin{equation}
\mathcal M_5 \sum_{\bm q}\langle\!\langle C_{\bm q\downarrow}S^+|C^\dagger_{\bm k'\uparrow}\rangle\!\rangle=\mathcal V_5\sum_{\bm q}\mathcal G_{\bm q\bm k'}+\mathcal W_5\sum_{\bm q}\mathcal F_{\bm q\bm k'},
\label{eq:D15}
\end{equation}
\begin{equation}
\mathcal M_6 \sum_{\bm q}\langle\!\langle C_{\bm q\uparrow}S^+|C^\dagger_{\bm k'\uparrow}\rangle\!\rangle=\mathcal V_6 \sum_{\bm q}\mathcal G_{\bm q\bm k'}+\mathcal W_6\sum_{\bm q}\mathcal F_{\bm q\bm k'}.
\label{eq:D16}
\end{equation}
}
To proceed, we perform the following sequence of substitutions: Eq.~(\ref{eq:D5}) into Eq.~(\ref{eq:D6}), Eq.~(\ref{eq:D4}) into Eq.~(\ref{eq:D3}), Eq.~(\ref{eq:D2}) into Eq.~(\ref{eq:D3}), Eq.~(\ref{eq:D3}) into Eq.~(\ref{eq:D1}), Eq.~(\ref{eq:D6}) into Eq.~(\ref{eq:D1}). Taking these substitutions together with the discussion above, we arrive at the following expressions

\begin{equation}
\mathcal M_{1}=1+2\left(n_g+\frac{2n_{h-} n_{h+}\left(n_g+\frac{n_{h-} n_{h+}}{1-n_g}\right)}{(1-n_g)(1+2n_g)+n_{h-} n_{h+}\left(1-\frac{n_g}{1-n_g}\right)}\right)\frac{n_g-\frac{n_{h-} n_{h+}}{1-n_g}}{1+n_g+\frac{n_{h-} n_{h+}}{1-n_g}},
\end{equation}
\begin{equation}
\mathcal V_1=S_{h-}+\frac{n_{h-}\left[2S_gn_g-K_g+\frac{n_{h-}}{1-n_g}\left(2S_g n_{h+}-K_{h+}\right)\right]}{(1-n_g)(1+2n_g)+n_{h-} n_{h+}\left(1-\frac{n_g}{1-n_g}\right)},
\end{equation}
\begin{equation}
\mathcal W_1=S_g+\frac{2n_{h-}S_{h+}\left(n_g+\frac{n_{h-}n_{h+}}{1-n_g}\right)}{(1-n_g)(1+2n_g)+n_{h-} n_{h+}\left(1-\frac{n_g}{1-n_g}\right)}-\left(n_g+\frac{2n_{h-}n_{h+}\left(n_g+\frac{n_{h-}n_{h+}}{1-n_g}\right)}{(1-n_g)(1+2n_g)+n_{h-} n_{h+}\left(1-\frac{n_g}{1-n_g}\right)}\right)\frac{K_g+\frac{n_{h+} K_b}{1-n_g}}{1+n_g+\frac{n_{h-} n_{h+}}{1-n_g}},
\end{equation}

\twocolumngrid

\begin{equation}
\mathcal M_2=1,
\end{equation}
\begin{equation}
\mathcal V_2=-n_{h+}\frac{\mathcal V_6}{\mathcal M_6}-n_g\frac{\mathcal V_3}{\mathcal M_3}-S_g,
\end{equation}
\begin{equation}
\mathcal W_2=-n_{h+}\frac{\mathcal W_6}{\mathcal M_6}-n_g\frac{\mathcal W_3}{\mathcal M_3}-S_{h+},
\end{equation}
\begin{equation}
\mathcal M_3=(1-n_g)(1+2n_g)+n_{h-} n_{h+}\left(1-\frac{n_g}{1-n_g}\right),
\end{equation}
\begin{equation*}
\mathcal V_3=2\frac{\mathcal V_6}{\mathcal M_6}n_{h+}\left(n_g+\frac{n_{h-}n_{h+}}{1-n_g}\right)+2S_g n_g-K_g
\end{equation*}
\begin{equation}
+\frac{n_{h-}}{1-n_g}\left(2S_g n_{h+}-K_{h+}\right),
\end{equation}
\begin{equation}
\mathcal W_3=2\frac{\mathcal W_6}{\mathcal M_6}n_{h+}\left(n_g+\frac{n_{h-}n_{h+}}{1-n_g}\right)+2S_{h+}\left(n_g+\frac{n_{h-}n_{h+}}{1-n_g}\right),
\end{equation}
\begin{equation}
\mathcal M_4=1-n_g,
\end{equation}
\begin{equation}
\mathcal V_4=n_{h+}\left(\frac{\mathcal V_3}{\mathcal M_3}+2\frac{\mathcal V_2}{\mathcal M_2}\right)+K_{h+},
\end{equation}

\begin{equation}
\mathcal W_4=n_{h+}\left(\frac{\mathcal W_3}{\mathcal M_3}+2\frac{\mathcal W_2}{\mathcal M_2}\right),
\end{equation}
\begin{equation}
\mathcal M_5=1-n_g,
\end{equation}
\begin{equation}
\mathcal V_5=n_{h-}\left(\frac{\mathcal V_6}{\mathcal M_6}-2\frac{\mathcal V_1}{\mathcal M_1}\right),
\end{equation}

\begin{equation}
\mathcal W_5=n_{h-}\left(\frac{\mathcal W_6}{\mathcal M_6}-2\frac{\mathcal W_1}{\mathcal M_1}\right)+ K_{h-},
\end{equation}
\begin{equation}
\mathcal M_6=1+n_g+\frac{n_{h-}n_{h+}}{1-n_g},
\end{equation}
\begin{equation}
\mathcal V_6=2\frac{\mathcal V_1}{\mathcal M_1}\left(\frac{n_{h-}n_{h+}}{1-n_g}-n_g\right),
\end{equation}
\begin{equation}
\mathcal W_6=2\frac{\mathcal W_1}{\mathcal M_1}\left(\frac{n_{h-}n_{h+}}{1-n_g}-n_g\right)-\left(\frac{n_{h+} K_{h-}}{1-n_g}+K_g\right).
\end{equation}
\onecolumngrid

\twocolumngrid

\bibliography{resistivity}

\end{document}